 \documentclass[preprint,12pt,authoryear]{elsarticle}

\usepackage{amssymb}  
\usepackage{ulem}
\usepackage{lipsum}
\usepackage[colorlinks = true, linkcolor = blue, urlcolor  = blue, citecolor = blue]{hyperref}
\usepackage{amsmath}
\usepackage{booktabs}
\usepackage{tabularx}

\newcommand{\pinocchio}{\texttt{PINOCCHIO}}
\newcommand{\CU}{\texttt{CU}}
\newcommand{\CUs}{\texttt{CUs}}
\newcommand{\kernel}{\texttt{kernel}}
\newcommand{\EDP}{\texttt{EDP}}
\newcommand{\GP}{\texttt{GP}}
\newcommand{\EDPs}{\texttt{EDPs}}

\journal{Astronomy $\&$ Computing}

\begin{document}

\begin{frontmatter}



\title{Accelerating cosmological simulations on GPUs: a step towards sustainability and green-awareness}


\author[OABR,ICSC]{G. Lacopo}

\author[U_TS,OATS,ICSC,IFPU]{M. D. Lepinzan}

\author[OATS,ICSC]{D. Goz}

\author[OATS,ICSC]{G. Taffoni}

\author[OATS,ICSC]{L. Tornatore}

\author[U_TS,OATS,ICSC,IFPUPM]{P. Monaco}

\author[PAWSEY]{P. J. Elahi}

\author[PAWSEY]{U. Varetto}

\author[PAWSEY]{M. Cytowski}

\author[IT4I]{L. Riha}

\affiliation[OABR]{organization={INAF-Osservatorio astronomico di Brera},
            addressline={Via E. Bianchi, 46}, 
            city={Merate},
            postcode={23807}, 
            state={LC},
            country={Italy}}
\affiliation[ICSC]{organization={ ICSC -- Centro Nazionale di Ricerca in High Performance Computing, Big Data e Quantum Computing},
            addressline={via Magnanelli 2}, 
            city={Bologna},
            postcode={40033}, 
            state={TS},
            country={Italy}}
\affiliation[U_TS]{organization={Dipartimento di Fisica, Sezione di Astronomia, Università di Trieste},
            addressline={via Tiepolo 11}, 
            city={Trieste},
            postcode={34143}, 
            state={TS},
            country={Italy}}
\affiliation[OATS]{organization={INAF-Osservatorio Astronomico di Trieste},
            addressline={via Tiepolo 11}, 
            city={Trieste},
            postcode={34143}, 
            state={TS},
            country={Italy}}
\affiliation[IFPU]{organization={IFPU -- Institute for Fundamental Physics of the Universe},
            addressline={via Beirut 2}, 
            city={Trieste},
            postcode={34151}, 
            state={TS},
            country={Italy}}
\affiliation[PAWSEY]{organization={Pawsey Supercomputing Centre}, 
            addressline={1 Bryce Avenue},
            city={Kensington WA},
            postcode={6151},
            country={Australia}}
\affiliation[IT4I]{organization={IT4Innovations National Supercomputing Center},
            addressline={Studentská 6231/1B}, 
            city={Ostrava},
            postcode={70800}, 
            country={Czech Republic}}

\begin{abstract}
The increasing complexity and scale of cosmological N-body simulations, driven by astronomical surveys like Euclid, call for a paradigm shift towards more sustainable and energy-efficient high-performance computing (HPC). The rising energy consumption of supercomputing facilities poses a significant environmental and financial challenge.

In this work, we build upon a recently developed GPU implementation of {\pinocchio}, a widely-used tool for the fast generation of dark matter (DM) halo catalogues, to investigate energy consumption. Using a different resource configuration, we confirmed the time-to-solution behavior observed in a companion study, and we use these runs to compare time-to-solution with energy-to-solution.

By profiling the code on various HPC platforms with a newly developed implementation of the Power Measurement Toolkit (PMT), we demonstrate an $8\times$ reduction in energy-to-solution and $8\times$ speed-up in time-to-solution compared to the CPU-only version. Taken together, these gains translate into an overall efficiency improvement of up to $64\times$. Our results show that the GPU-accelerated \pinocchio\ not only achieves substantial speed-up, making the generation of large-scale mock catalogues more tractable, but also significantly reduces the energy footprint of the simulations. This work represents an step towards ``green-aware" scientific computing in cosmology, proving that performance and sustainability can be simultaneously achieved.
\end{abstract}




\end{frontmatter}




\section{Introduction}
Over the last decades, the cosmological community has dedicated substantial effort to increase the level of detail in simulations, aiming to reconstruct the birth and growth of cosmic structures and to shed light on the nature of the dark sector of the Universe with unprecedented accuracy. To achieve this, simulations are nowadays run with up to trillions of particles, leading to memory requirements of hundreds of terabytes or even petabytes. 

Meeting such demands requires high efficiency, and maximizing the potential of modern High Performance Computing (HPC) platforms, particularly through Graphics Processing Unit (GPU) acceleration, has therefore become imperative. Consequently, cosmological simulations need to be run on pre-exascale and exascale supercomputing platforms, many of which are currently ranked in Top500, HPCG500 and Green500 lists\footnote{\url{https://top500.org/}}. 

The Top500 list ranks the HPC systems according to their rank in the HPL (High Performance Linpack), a benchmark which solves linear systems of equations with an exceptional level of optimizations. However, the HPL is poorly representative for actual scientific simulations, as the impact of memory transfer is largely ignored. To address this limitation, a new benchmark named HPCG (High Performance Conjugate Gradients) has been introduced to test this systems under more realistic conditions, better reflecting the behavior of real scientific applications. When running under full workload, this HPC systems show a power absorption on the order of tens of $MW$ (megawatts).

Large cosmological simulations require hundreds of thousands core/node hours, forcing HPC facilities to operate at full capacity for extended periods. Sustaining such workloads for such  amount of time leads to unsustainable energy costs, especially in an age of environmental awareness \citep[see e.g.][]{space2,10.1117/12.3020361}. It is easy to understand that in the next future the number of HPC facilities will increase and their power demands will rise accordingly, potentially becoming unsustainable. For this reason, in the last decades ``green awareness" has become increasingly important, and nowadays pure performance is not enough to establish the efficiency of a scientific code. In particular, hardware and software development must advance at the same pace, and co-design is required to guarantee that scientific codes runs efficiently on modern HPC facilities~\citep{10.1007/978-3-030-22871-2_14,10.1007/978-3-030-32520-6_33,computation8020034, lacopo2025hpc}.

This work focuses on evaluating the energy efficiency of the {\pinocchio} code~\citep{Monaco:2001jg, Monaco:2013qta}, a framework designed to simulate the formation and growth of DM halos. Unlike full N-body simulations, which require computing the gravitational interaction between all particles, {\pinocchio} provides a much faster alternative by relying on Lagrangian Perturbation Theory (LPT). Thanks to this feature, and to its parallel CPU implementation based on MPI and OpenMP, {\pinocchio} plays a crucial role in generating mock catalogs of DM halos for large-scale surveys such as \textit{Euclid}~\citep{Euclid:2024yrr}, where both speed and scalability are essential to produce thousands realizations of the Universe~\citep{Euclid:2025lfa} required for robust statistical analyses, including covariance matrix estimation and calibration~\citep{Euclid:2021api,Euclid:2022txd, Salvalaggio:2024vmx}.

In a companion work~\citep{Lepinzan_in_prep}, an embarrassingly parallel module of the code was offloaded to GPUs using OpenMP~\citep{OpenMP} target directives, improving performance and portability across architectures such as NVIDIA and AMD. Building on that effort, the present work evaluates the energy impact of this GPU porting by measuring and comparing the energy consumption of CPUs and GPUs on different architectures. For this purpose, we integrate the Power Measurement Toolkit (PMT)~\citep{PMT}, a high-level library to monitor power consumption between CPUs and GPUs, allowing us to assess the energy efficiency of the optimized code.
 
Through comparative benchmarking on two major supercomputing platforms, KAROLINA (NVIDIA-based) and SETONIX (AMD-based), we demonstrate both the portability and performance gains of our approach, achieving a speedup of about $2\times$ on the NVIDIA system and at least $8\times$ on the AMD, the latter benefiting from better architectural features detailed in Section~\ref{GPU_architectures}. These results are consistent with those reported in~\cite{Lepinzan_in_prep}, although in the present work tests are performed using the entire node as the compute unit, rather than a quarter of the node. In addition, GPU offloading yields up to an $8\times$ reduction in energy consumption, leading to an overall efficiency improvement of up to $64\times$ on the AMD platform, and about $4\times$ on the NVIDIA system, again reflecting architectural differences. Taken together, these results highlight the potential of OpenMP-based GPU offloading to accelerate scientific simulations efficiently across diverse hardware accelerators. 

This work is organized as follows. Section~\ref{Computing_platforms} presents the key characteristics of the computing platforms used in this study. Section~\ref{Pinocchio_GPU_PMT} briefly introduces the main components of the {\pinocchio} code, emphasizing the embarrassingly parallel nature of the target segment~\citep[a more detailed description is provided in][]{Lepinzan_in_prep}. Section~\ref{parallel-pmt} describes the PMT library and its new parallel implementation for MPI-based codes such as {\pinocchio}. Section~\ref{Methodology} defines the metrics adopted to evaluate efficiency and outlines the resource configurations used for both strong- and weak-scaling tests. Section~\ref{Results} presents the results of our energy-efficiency measurements on both NVIDIA- and AMD-based platforms. Finally, conclusions are given in Section~\ref{sec:conclusions}.

\section{Computing platforms}
\label{Computing_platforms}

To assess the portability, performance and energy efficiency of our GPU porting strategy, we conducted benchmarks on two state-of-the-art supercomputing platforms, described in detail below. These systems, KAROLINA and SETONIX, represent distinct GPU architectures. The selection of these platforms is driven by two key considerations: i) the requirement for access to energy counters, which are commonly unavailable to general users on HPCG500 supercomputing platforms, and ii) leveraging the architectural difference (NVIDIA on KAROLINA, AMD on SETONIX) to address the core objective of this paper and its associated work \citep{Lepinzan_in_prep}: code portability. These two HPC platforms offer an ideal environment to evaluate cross-platform compatibility, scalability, and energy-aware performance. While we currently focus on the two main GPU vendors, future plans include an extended analysis to incorporate new clusters featuring Intel GPUs.

\subsection{NVIDIA cluster}
KAROLINA is part of the IT4I \footnote{\url{https://www.it4i.cz/en/infrastructure/karolina}} Czech national HPC facility, ranked as $66^{th}$ in the November 2025 HPCG500 list and $84^{th}$ in the November 2025 Green500 list. The platform consists of 72, 64-cores, 2 $\times$ AMD EPYC 7H12 CPU nodes equipped with 8 NVIDIA Tesla Ampere 100 GPUs. Each CPU node has 1 TB of DDR4 memory, while each GPU provides 40 GB of HBM2 memory. The CPU and the GPU are interfaced by a PCIe Gen4 interconnection. Computing nodes are connected through a Mellanox HDR Infiniband network with DragonFly+ topology, with 200 GB/s bidirectional bandwidth. 

The {\pinocchio} code was compiled on this system using the NVC/NVC++ compilers v24.3 and OpenMPI library v5.1 was employed.

\subsection{AMD cluster}
The Setonix-GPU partition, is a constituent component of the HPC infrastructure at the Pawsey Supercomputing Centre in Perth\footnote{\url{https://pawsey.org.au/}}, Western Australia, ranked as $52^{th}$ in the November 2025 HPCG500 list and $30^{th}$ in the November 2025 Green500 list.
The partition consists of 154 compute nodes, each featuring a dual-device configuration. Each node integrates one AMD-optimized 3rd Generation EPYC ``Trento" processor, providing 64 cores, which is coupled with 256 GB of host memory (CPU). The accelerator complex consists of four AMD Instinct MI250X accelerators, collectively delivering eight Graphics Compute Dies (GCDs) (two GCDs per accelerator), complemented by 128 GB of High Bandwidth Memory 2e (HBM2e).

Intra-node communication, encompassing both CPU-GPU and GPU-GPU interconnects, is facilitated by the AMD Infinity Fabric technology. Inter-node connectivity is established via a HPE Slingshot-11 network fabric, leveraging a DragonFly+ topology to achieve a 200 GB/s bidirectional bandwidth between distinct computing nodes.

The {\pinocchio} code was compiled on this system with the amdclang-18 compiler and MPICH-8.1.31 implementation was employed.

\subsection{Architectural Comparison and Performance Implications}
\label{GPU_architectures}
To better understand the performance differences observed in our benchmarks, we provide a detailed analysis of the GPU architectures and their interconnection topologies on both platforms, with the primary specifications of the NVIDIA A100 (Ampere) and AMD MI250X (CDNA2) accelerators summarized in Table~\ref{tab:hardware_specs}.

\begin{table*}[h!]
    \centering
    \caption{Key Hardware Specifications of the two accelerators. For a detailed description of the computing node architecture of the two HPC clusters, see \ref{architecture_appendix}.}
    \label{tab:hardware_specs}
    
    \setlength{\extrarowheight}{1pt} 

    \begin{tabularx}{\textwidth}{l c c}
        \toprule
        \textbf{Feature} & \textbf{NVIDIA A100} & \textbf{AMD MI250X} \\
        \midrule
        Process Node & 7\,nm & 6\,nm (MCM) \\
        Compute Architecture & Ampere & CDNA2 \\
        \midrule
        \textbf{Compute Units} & & \\
        Units / Dies & 108 SMs & 2 $\times$ 110 CUs \\
        FP64 Cores / SPs & 6,912 & 2 $\times$ 7,040 \\
        \midrule
        \textbf{Performance (FP64)} & & \\
        Peak Theoretical & 9.7 TFLOPS & 47.9 TFLOPS \\
        \midrule
        \textbf{Memory Subsystem} & & \\
        Total HBM & 40 GB HBM2 & 128 GB HBM2e \\
        Total Bandwidth & 1,555 GB/s & 3,276 GB/s \\
        \midrule
        \textbf{Cache Hierarchy} & & \\
        L2 Cache & 40 MB & 2 $\times$ 8 MB \\
        L1 / Local Cache & 192 KB per SM & 16 KB per CU \\
        \midrule
        \textbf{Thermal Design Power (TDP)} & & \\
        Power & 400 W & 560 W\\
        \midrule
        \textbf{CPU-GPU interconnection} & & \\
        Connection & PCIe 4.0 & Infinity Fabric 3 \\
        Bidirectional Bandwidth & 31.5 GB/s & 36-50 GB/s \\
        \bottomrule
    \end{tabularx}
    \setlength{\extrarowheight}{0pt}
\end{table*}

The approximately $2.5\times$ difference in FP64 peak performance (23.9 vs 9.7 TFLOPS) between AMD GCDs and NVIDIA A100s and the difference in bidirectional bandwidth for CPU-GPU interconnection largely explain the superior speed-ups observed on the AMD platform for a compute-bound kernel as discussed in Section~\ref{Results}.

The combination of superior FP64 performance, better host-device integration, and the kernel compute-bound nature allows the AMD to exploit its architectural advantages more effectively. As a results,  our benchmarks show AMD achieving roughly $4\times$ the speedup of NVIDIA ($8\times$ vs $2\times$ relative to CPU baseline). A more detailed discussion of these differences and their impact is provided in Section~\ref{Results}.

Overall this architectural analysis highlights the value of portable programming models such as OpenMP, which enable scientific applications to exploit the strengths of different GPU architectures without relying on platform-specific tuning.

\subsubsection{Power Efficiency Considerations}

While the AMD MI250X has a higher Thermal Design Power (TDP) of 560W compared to the NVIDIA A100's 400W, the power efficiency story is more nuanced when considering FP64 performance per watt. The MI250X delivers approximately 85.5 GFLOPS/W (47.9 TFLOPS at 560W per card, considering both GCDs), while the A100 provides 24.3 GFLOPS/W (9.7 TFLOPS at 400W), resulting in a $3.5\times$ advantage in raw FP64 efficiency for AMD. This theoretical advantage is validated by the two platforms positions in the Green500 ranking, with the SETONIX cluster being ahead of KAROLINA in the list. 

For our compute-bound kernel achieving $8\times$ speedup on AMD versus $2\times$ on NVIDIA (compared to CPU baseline), we can theoretically expect the AMD platform to complete the same workload using approximately $(1.4 \times \text{power}) \times (0.5 \times \text{time}) = 0.7\times$ the energy of the NVIDIA platform, representing a 30\% energy saving despite the higher instantaneous power draw. However, this efficiency advantage must be weighed against practical considerations including cooling infrastructure requirements and the higher power delivery demands of the MI250X.
 
\section{{\pinocchio} code}
\label{Pinocchio_GPU_PMT}
{\pinocchio}~\citep[\texttt{PIN}pointing \texttt{O}rbit \texttt{C}rossing-\texttt{C}ollapsed \texttt{HI}erarchical \texttt{O}bjects][]{Monaco:2001jg,Monaco:2013qta,Munari:2016aut}, is a fast, massively parallel code based on LPT that simulates the distribution of DM halos from an initial density field. The code proceeds in three stages:

\begin{enumerate}
    \item Generation of a linear density contrast field $\delta$ on a uniform grid; 
    \item Calculation of a ``collapse time'' for each grid point;
    \item Assembly of collapsed particles into structures, such as halos or filaments, by tracing their hierarchical formation.
\end{enumerate}

A detailed description of the algorithmic stages of {\pinocchio} is provided in~\cite{Euclid:2025lfa} \citep[see also][]{Lepinzan_in_prep}. The GPU porting effort builds on the legacy version of the code, publicly available at \url{https://github.com/pigimonaco/Pinocchio.git}
, which was originally designed for multi-core CPU architectures using a hybrid MPI/OpenMP approach~\citep{Monaco:2013qta}. 

The part of the code offloaded to GPUs, presented in~\cite{Lepinzan_in_prep}, corresponds to the second step of {\pinocchio}, namely the computation of collapse times at each grid point. This operation is inherently independent across grid points, making it embarrassingly parallel and therefore well suited for GPU offloading. 

In the present work, we evaluate the energy efficiency of this stage across different architectures. For clarity, throughout the paper we will refer to this collapse time computation simply as the \textit{kernel}.

\section{Parallel PMT}
\label{parallel-pmt}
Power Measurements Toolkit~\footnote{\url{https://git.astron.nl/RD/pmt.git}}~\citep{PMT} is a C++ library that allows energetic profiling of code portions in a plethora of architectures, such as Intel and AMD CPUs through Running Average Power Limit (RAPL) counters~\citep{david2010rapl}, NVIDIA GPUs through Nvidia Management Library (NVML)~\citep{kasichayanula2012power}, and AMD GPUs through Radeon Open Compute platform (ROCM)~\footnote{\url{https://www.amd.com/en/products/software/rocm.html}} and AMD System Management Interface (SMI)\footnote{\url{https://rocm.docs.amd.com/projects/amdsmi/en/latest/}}. For the sake of completeness, PMT also provides an opportunity to measure the energy for other architectures, like Xilinx Field-Programmable Gate Arrays (FPGAs).

PMT is perfectly suitable for serial codes, when the main process profiles the relevant \textit{kernel} or code portion and outputs the resulting energy, runtime, and power. However, the situation is more complex for MPI codes, such as \pinocchio. Allowing every process to read hardware counters and output their own results would be redundant and inefficient. To address this limitation, we developed a parallel version of PMT that collects the results pertaining to all MPI processes and writes a final report for all CPUs/GPUs profiling. 

The Parallel PMT is a C++ library with a C wrapper, making it usable within any C/C++ application. It relies on the standard PMT library and therefore cannot be compiled unless PMT is already available on the system. The current implementation supports RAPL counters for CPUs and provides a flag to specify whether the GPUs (if present) are NVIDIA or AMD. A detailed description of how to use the library in scientific applications is provided in \ref{pmt_appendix}.

This parallel version of the PMT library is expected to be released in the near future. Unless otherwise specified, all energy measurements presented in this work were obtained using this specific version.

\section{Methodology}
\label{Methodology}

From energy-to-solution and time-to-solution considered separately, it is possible to infer which configuration is the greenest and which is the fastest. However, it is more informative to combine these metrics into a single quantity that gives information about the most efficient configuration, where for efficient we refer to the one which minimizes the combination of energy-to-solution and time-to-solution.
Hereafter, the term \textit{configuration} will be assigned to a specific CPU or GPU run with $N$ compute units.

In this section, we first describe the resource configurations adopted for the benchmarks, then we introduce the two metrics used to evaluate efficiency: the Energy-Delay Product (\EDP)~\citep{computation8020034, 10.1007/978-3-030-32520-6_33} and the Green Productivity (\GP)~\citep{lacopo2025greencomputingskaera,10974809}.

\subsection{Benchmarks}
\label{benchmarks}

To ensure a meaningful comparison across the different computing platforms, we perform the benchmarks by choosing as a computational unit (hereafter {\CU}) the compute node. 

The choice is motivated by the fact that, for most applications, the primary interest lies in the energy which is consumed by the entire node. Otherwise, non-trivial models would be required to estimate the idle consumption of non-active components (CPUs or GPUs), which must still be accounted for. In pure CPU runs, the idle GPU consumption is included automatically since each MPI task activates PMT GPU profiling to account for the imprint of accelerators. Finally, since the parallel PMT library relies on MPI, it is also relevant to test its behavior across multiple computing nodes. 

For our specific purpose, we adopted the following configurations. These choices are motivated by the heterogeneity of the target architectures, as detailed below:
\begin{itemize}
    \item On KAROLINA, each MPI process can manage one GPU out of eight using at most sixteen cores (i.e. spawning sixteen OMP threads). The {\CU} will be the node, with 8 MPI tasks, 16 OpenMP threads and one GPU per task;
    \item On SETONIX, each chiplet, to which MPI processes are binded, consists of eight cores and is associated with one GPU. The {\CU} will again be the node, with 8 MPI tasks, 8 OpenMP threads and one GPU per task.
\end{itemize}

The PMT library does not account for the energy consumed by inter-node communication. This omission does not impact our treatment since the {\kernel} computation is embarrassingly parallel and no MPI communication is needed during its execution.

In all runs, the {\kernel} time-to-solution is the wall-clock time required by the actual computation, as measured by the slowest MPI task. The energy-to-solution is always the one read by the highest energy demanding MPI task. For GPU runs, both time-to-solution and energy-to-solution include the cost of pure GPU computation as well as host–device data transfers.

\subsection{Strong scaling configurations}
\label{strong_scaling_config}
Strong scaling measures how the execution time of a fixed-size problem changes as the number of processors (or computing elements) is increased.
In this specific case, we keep the number of {\pinocchio} particles fixed while increasing the {\CUs}~at each step. In these tests we use up to 16 {\CUs}. For both CPU and GPU runs, we set the {\pinocchio} number of particles fixed to $768^3$. Configurations for the SETONIX cluster are summarized in Table \ref{table:strong-scaling-table}.

\begin{table}[t]
\begin{center}
\centering \tabcolsep 3pt
\resizebox{0.70\columnwidth}{!}{%
\begin{tabular}{l|c|c|c|c|c|c|c|c}
\hline
& Nodes & MPI(threads) & GPUs & Grid size \\ 
\hline
CPU & 1 & 8 (8) & 0 & $768^3$ \\
\hline
CPU & 2 & 16 (8) & 0 & $768^3$ \\
\hline
CPU & 4 & 32 (8) & 0 & $768^3$ \\
\hline
CPU & 6 & 48 (8) & 0 & $768^3$ \\
\hline
CPU & 8 & 64 (8) & 0 & $768^3$ \\
\hline
CPU & 12 & 96 (8) & 0 & $768^3$ \\
\hline
CPU & 16 & 128 (8) & 0 & $768^3$ \\
\hline
GPU & 1 & 8 (8) & 8 & $768^3$ \\
\hline
GPU & 2 & 16 (8) & 16 & $768^3$ \\
\hline
GPU & 4 & 32 (8) & 32 & $768^3$ \\
\hline
GPU & 6 & 48 (8) & 48 & $768^3$ \\
\hline
GPU & 8 & 64 (8) & 64 & $768^3$ \\
\hline
GPU & 12 & 96 (8) & 96 & $768^3$ \\
\hline
GPU & 16 & 128 (8) & 128 & $768^3$ \\
\hline
\end{tabular}
}
\end{center}
\caption{SETONIX configurations for CPU only and GPU runs adopted for \textit{strong} scaling tests. The first column indicates the number of nodes. The second column reports to the total number of MPI tasks and the number of threads spawned by each task. The third column specifies the number of GPUs, while the last column gives the fixed box size adopted in the simulations. For GPU runs, the number of MPI tasks corresponds to the number of GPUs.}
\label{table:strong-scaling-table}
\end{table}

Since the number of CPU cores differs from SETONIX to KAROLINA, which is constituted by dual-socket nodes, we also summarize the configuration for the latter cluster in Table \ref{table:strong-scaling-table-nvidia}.

\subsection{Weak scaling configurations}
\label{weak_scaling_config}
Weak scaling measures how the execution time changes as both the number of processors and the size of the problem (workload per processor) are increased proportionally in order to maintain a fixed amount of work per processor.
In this specific case, we double both the total number of particles and \CUs~at each step, ranging from $512^3$ up to $1024^3$. The tests started from the 1 \CU~configuration and scaled up to 8 {\CUs}. All configurations, together with the corresponding particle counts, are summarized in Table~\ref{table:weak-scaling-table} and Table~\ref{table:weak-scaling-table-nvidia}, for the SETONIX and KAROLINA clusters, respectively.
 
\begin{table}[t]
\begin{center}
\centering \tabcolsep 3pt
\resizebox{0.70\columnwidth}{!}{%
\begin{tabular}{l|c|c|c|c|c|c|c|c}
\hline
& Nodes & MPI(threads) & GPUs & Grid size \\ 
\hline
CPU & 1 & 8 (16) & 0 & $768^3$ \\
\hline
CPU & 2 & 16 (16) & 0 & $768^3$ \\
\hline
CPU & 4 & 32 (16) & 0 & $768^3$ \\
\hline
CPU & 6 & 48 (16) & 0 & $768^3$ \\
\hline
CPU & 8 & 64 (16) & 0 & $768^3$ \\
\hline
CPU & 12 & 96 (16) & 0 & $768^3$ \\
\hline
CPU & 16 & 128 (16) & 0 & $768^3$ \\
\hline
GPU & 1 & 8 (16) & 8 & $768^3$ \\
\hline
GPU & 2 & 16 (16) & 16 & $768^3$ \\
\hline
GPU & 4 & 32 (16) & 32 & $768^3$ \\
\hline
GPU & 6 & 48 (16) & 48 & $768^3$ \\
\hline
GPU & 8 & 64 (16) & 64 & $768^3$ \\
\hline
GPU & 12 & 96 (16) & 96 & $768^3$ \\
\hline
GPU & 16 & 128 (16) & 128 & $768^3$ \\
\hline
\end{tabular}
}
\end{center}
\caption{Same as Table~\ref{table:strong-scaling-table} but for KAROLINA.}
\label{table:strong-scaling-table-nvidia}
\end{table}

\begin{table}[h!]
\begin{center}
\centering \tabcolsep 3pt
\resizebox{0.70\columnwidth}{!}{%
\begin{tabular}{l|c|c|c|c|c|c|c|c}
\hline
& Nodes & MPI(threads) & GPUs & Grid size  \\ 
\hline
CPU & 1 & 8 (8) & 0 & $512^3$ \\
\hline
CPU & 2 & 16 (8) & 0 & $644^3$ \\
\hline
CPU & 4 & 32 (8) & 0 & $812^3$ \\
\hline
CPU & 8 & 64 (8) & 0 & $1024^3$ \\
\hline
GPU & 1 & 8 (8) & 8 & $512^3$ \\
\hline
GPU & 2 & 16 (8) & 16 & $644^3$ \\
\hline
GPU & 4 & 32 (8) & 32 & $812^3$ \\
\hline
GPU & 8 & 64 (8) & 64 & $1024^3$ \\
\hline
\end{tabular}
}
\end{center}
\caption{SETONIX configurations for CPU-only and GPU runs adopted for \textit{weak} scaling tests. The column structure is the same as in Table~\ref{table:strong-scaling-table}.}
\label{table:weak-scaling-table}
\end{table}

\begin{table}[t]
\begin{center}
\centering \tabcolsep 3pt
\resizebox{0.70\columnwidth}{!}{%
\begin{tabular}{l|c|c|c|c|c|c|c|c}
\hline
& Nodes & MPI(threads) & GPUs & Grid size  \\ 
\hline
CPU & 1 & 8 (16) & 0 & $512^3$ \\
\hline
CPU & 2 & 16 (16) & 0 & $644^3$ \\
\hline
CPU & 4 & 32 (16) & 0 & $812^3$ \\
\hline
CPU & 8 & 64 (16) & 0 & $1024^3$ \\
\hline
GPU & 1 & 8 (16) & 8 & $512^3$ \\
\hline
GPU & 2 & 16 (16) & 16 & $644^3$ \\
\hline
GPU & 4 & 32 (16) & 32 & $812^3$ \\
\hline
GPU & 8 & 64 (16) & 64 & $1024^3$ \\
\hline
\end{tabular}
}
\end{center}
\caption{Same as Table~\ref{table:weak-scaling-table} but for KAROLINA.}
\label{table:weak-scaling-table-nvidia}
\end{table}

\subsection{Energy-Delay Product}
\label{edp-section}
The {\EDP} is defined as the product of the energy consumed by a specific configuration and the corresponding time-to-solution, as defined below:
\begin{equation}
    \label{edp-equation}
    EDP = E \times T^w
\end{equation}
where $w$ is a weight factor which determines the importance given to time-to-solution compared to energy-to-solution. In this work we present our results with $w = 1,2,3$. The configuration which minimizes the {\EDP} is the most efficient one.

\subsection{Green Productivity}
\label{green-prod}
By examining the {\EDP} for different configurations, it is possible to identify which one has the lowest \EDP. However, it is also useful to define a metric that relates a given configuration to a reference one, in order to asses which is the most efficient. This motivates the introduction of the green productivity, \GP, defined as follows:
\begin{equation}
    \label{gp-equation}
    GP = \frac{T_0/T_N}{\alpha E_N/E_0}
\end{equation}
where $T_0$ and $E_0$ are time-to-solution and energy-to-solution of the reference configuration, respectively, and $T_N$ and $E_N$ are the corresponding quantities for the tested configuration. The parameter $\alpha$ specifies the relative weight assigned to energy-to-solution compared to time-to-solution. 

From this definition, it follows that when the same code implementation is tested by varying only the number of {\CUs}, the numerator reduces to the runtime speedup, while the denominator corresponds to the inverse of energy speedup. For \GP~analyses, we assign equal importance the two quantities by setting $\alpha=1$. In this way, the configuration that maximizes \GP~is the most efficient one.

\section{Results}
\label{Results}
In this Section we present the results of strong- and weak-scaling tests, with a focus on the scalability of energy-to-solution. Since in many cases the combination of energy-to-solution and time-to-solution is more informative, we also report the \EDP\ (Section~\ref{edp-section}) and \GP\ (Section~\ref{green-prod}) results.
Results obtained on the SETONIX cluster are presented in~\ref{setonix-res}, while those for the KAROLINA cluster are given in Section~\ref{karolina-res}. To understand which architecture is more efficient to run our \kernel, in both CPU and GPU configurations, we present a comparison between the \EDPs~of the two clusters in Section~\ref{sp:norm_edp}. It is important to stress that this comparison is machine-dependent, and cannot be generalized to all NVIDIA and AMD architectures, since machine topology, CPU-GPU interconnection and the number of CPU cores per GPU impact on the results.

\subsection{Setonix cluster}
\label{setonix-res}

\subsubsection{Strong scaling tests}
\label{strong-scaling}

\begin{figure}[t]
    \centering
    \includegraphics[width=1.0\textwidth]{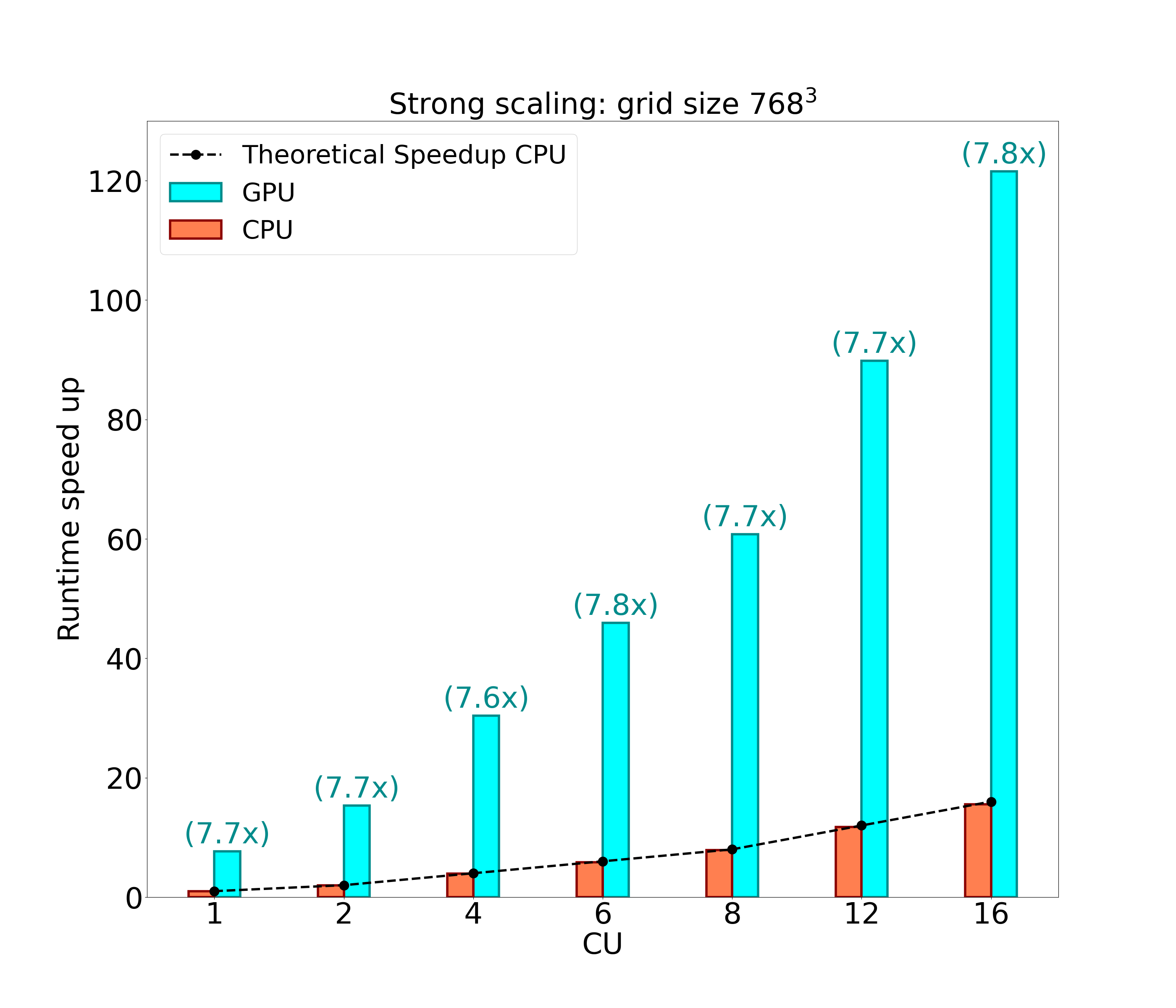}
    \caption{Strong scaling runtime speedup for the SETONIX cluster. All the histograms are normalized to the 1 CPU \CU~reference. The numbers above the GPU bar indicate the gain factor compared to the same \CU~CPU configuration.}
    \label{Runtime-strong}
\end{figure}

The runtime strong scaling results for the SETONIX cluster are shown in Figure~\ref{Runtime-strong}. All histograms are normalized to the 1 CPU \CU~configuration, so their height represents the relative increase with respect to the reference. The numbers above each GPU bar indicate the speedup of GPUs compared to CPUs with exactly the same number of \CUs. Across all runs, GPU runtime gain remains stable around $7.7-7.8\times$ times over the corresponding CPU configuration, consistent with those reported in~\cite{Lepinzan_in_prep} .

\begin{figure}[h]
    \centering
    \includegraphics[width=1.0\textwidth]{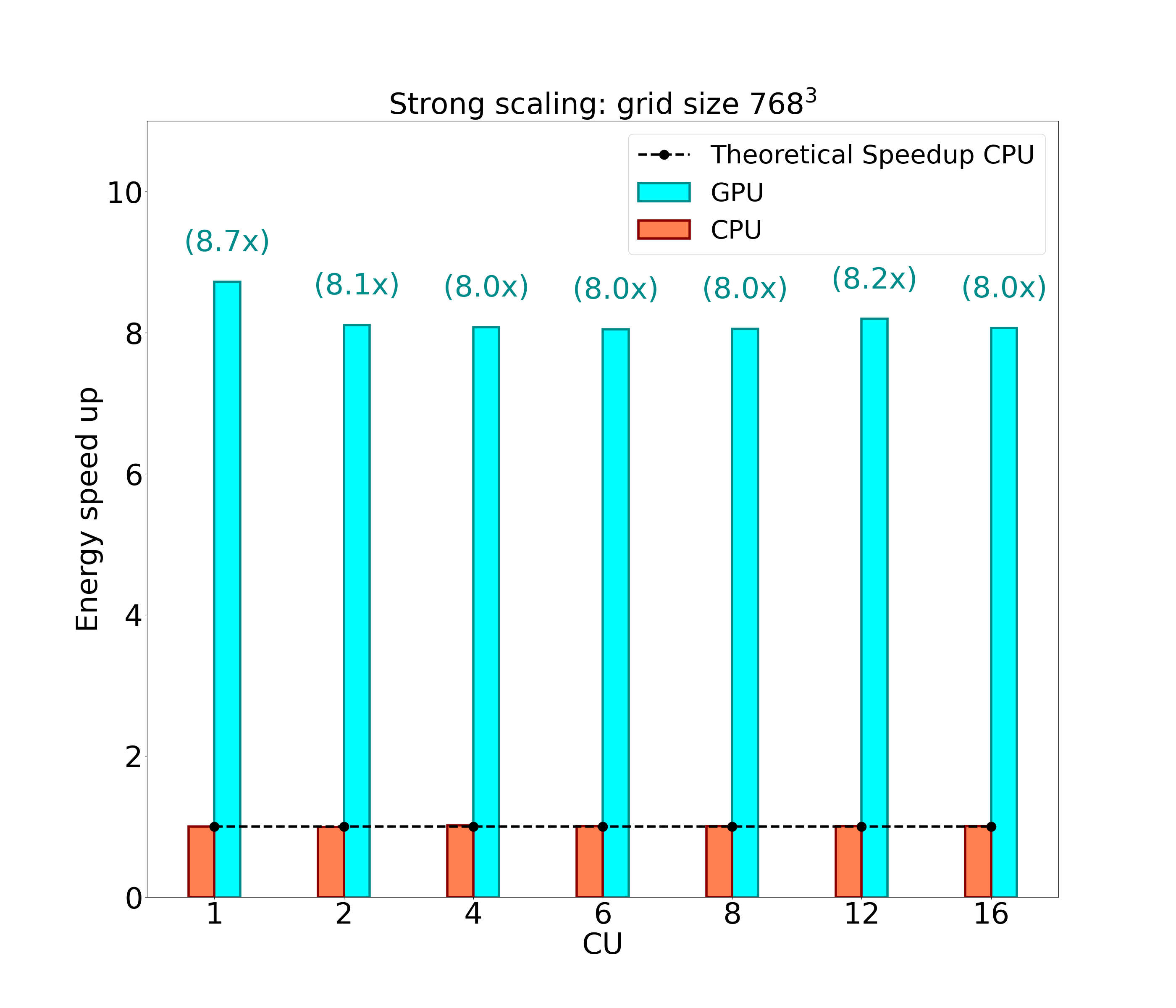}
    \caption{Strong scaling energy speedup for the SETONIX cluster.  All histograms are normalized and annotated as in Figure~\ref{Runtime-strong}.}
    \label{Energy-strong}
    
\end{figure}

Figure \ref{Energy-strong} shows the energy speedup results on the SETONIX cluster, where all histograms are normalized to the reference configuration (1 CPU \CU). The speedup is nearly flat because the problem (box) size is fixed while the computing resources increase at each step. For an embarrassingly parallel code, doubling the {\CUs} reduces the runtime, as shown in Figure~\ref{Runtime-strong}, but the energy remains constant, since the decrease in runtime is balanced by the increase in allocated resources. Again, the numbers above the GPU histograms indicate the speedup relative to the corresponding CPU configuration. Overall, the gain in energy-to-solution from using GPUs exceeds $8\times$.

\begin{figure}[h]
    \centering
    \includegraphics[width=1\textwidth]{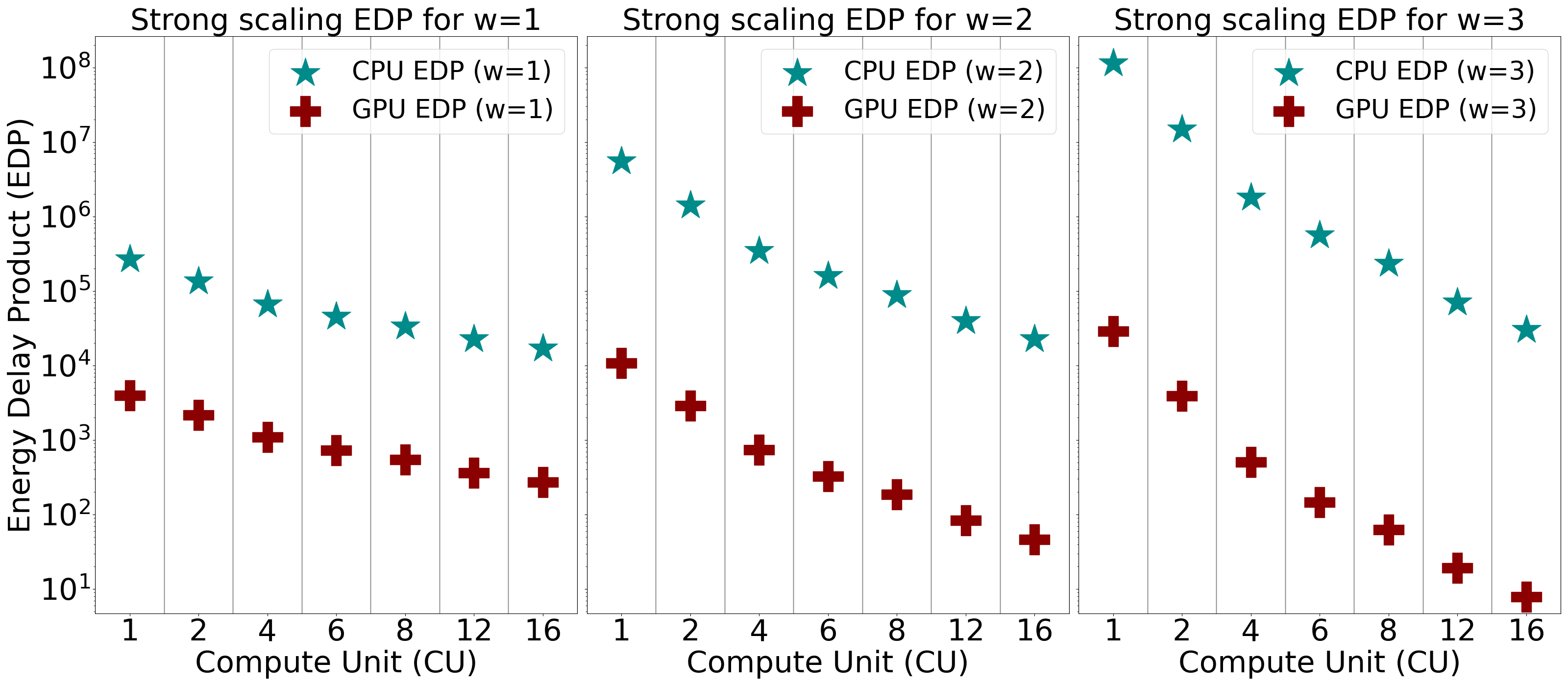}
    \caption{\EDP~for the SETONIX cluster in strong scaling tests, for all CPU and GPU configurations. Results are shown for $w = 1,2,3$.}
    \label{EDP-strong}
\end{figure}

Figure \ref{EDP-strong} shows the \EDP~results in strong scaling tests on the SETONIX cluster, with weight factors $w = 1,2,3$. These tests were performed for all configurations reported in Table~\ref{table:strong-scaling-table}.  Cyan stars denote pure CPU runs, while dark red Greek crosses correspond to GPU runs. The y-axis is in logarithmic scale. For a fixed number \CUs, even when $w=1$, GPUs are more efficient than their CPU counterpart by almost a factor of $\sim 64\times$. This is consistent with Figures~\ref{Runtime-strong} and \ref{Energy-strong}, which show that GPUs are both faster and greener by a factor of $\sim 8\times$. 

The \EDP~plot (Figure \ref{EDP-strong}) shows the dramatic advantage of GPUs when combining energy-to-solution and time-to-solution. When a greater emphasis is placed on runtime gains (i.e. $w = 2,3$), the gap between CPU and GPU runs becomes increasingly pronounced. The \EDP~decreases as a function of \CUs, indicating that employing more resources in these strong-scaling tests improves system utilization: energy consumption remains constant, while runtime halves at each step.

\subsubsection{Weak scaling tests}
\label{weak-scaling}

\begin{figure}[t]
    \centering
    \includegraphics[width=1.0\textwidth]{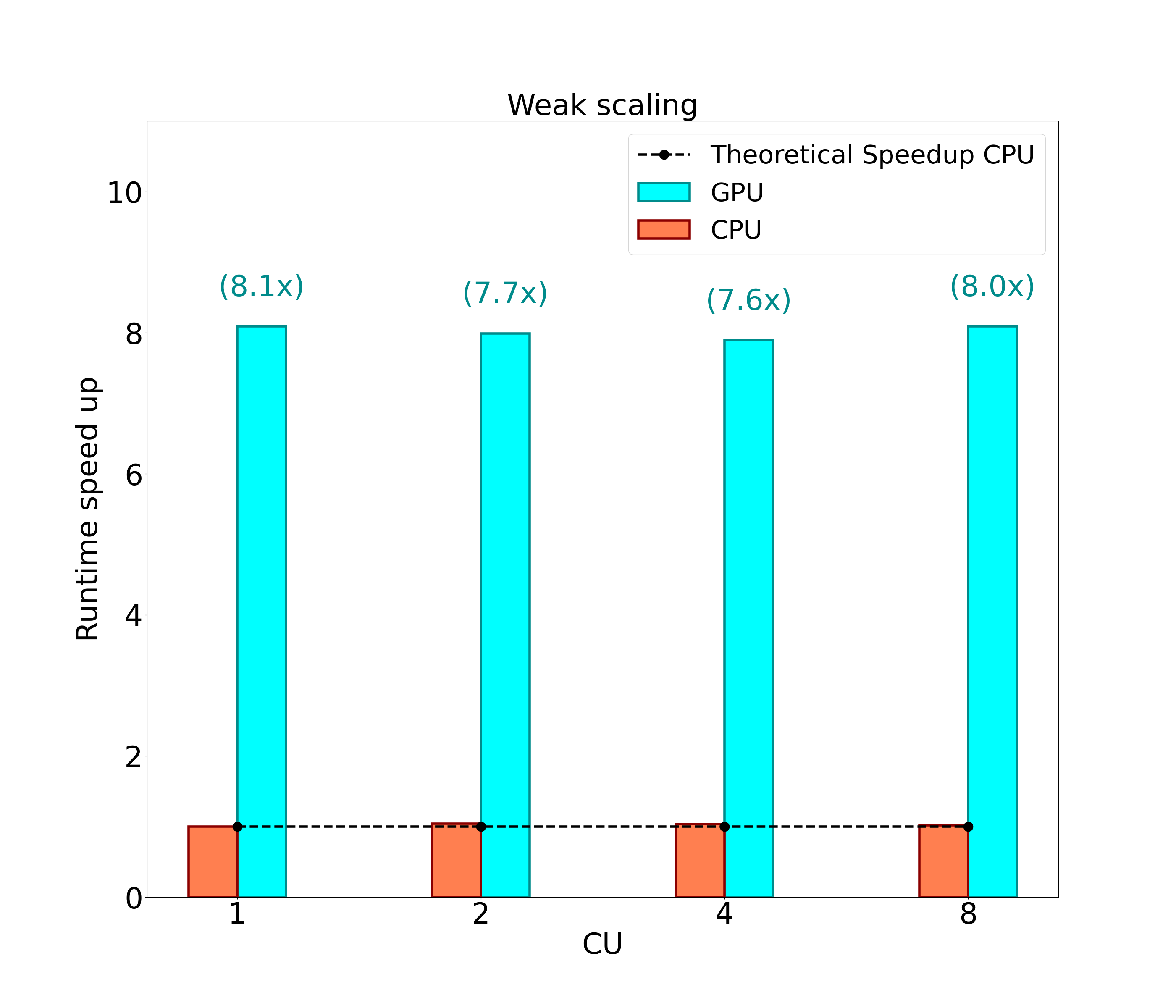}
    \caption{Same as Figure~\ref{Runtime-strong}, but for the weak scaling.}
    \label{Runtime-weak}
\end{figure}

\begin{figure}[h]
    \centering
    \includegraphics[width=1.0\textwidth]{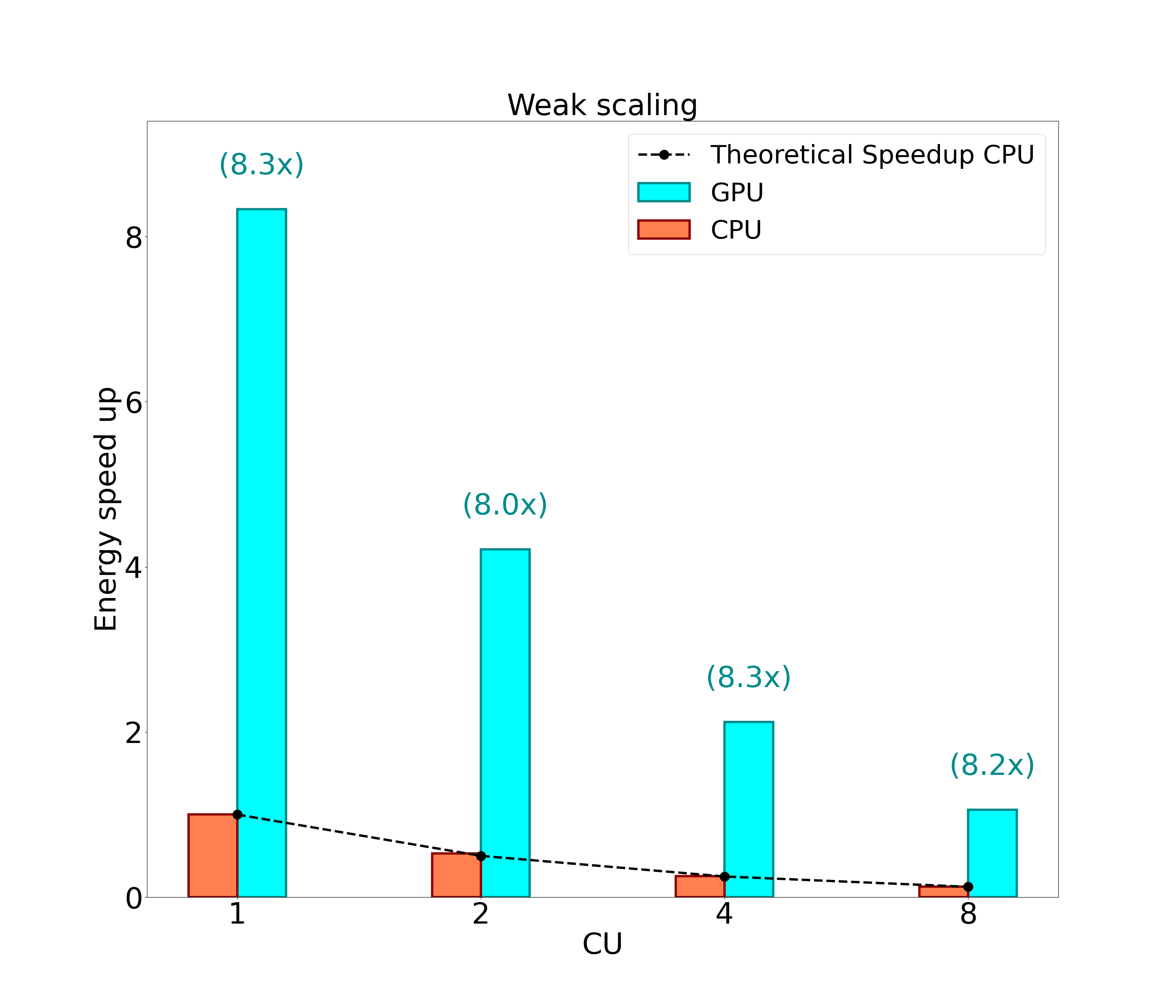}
    \caption{Same as Figure~\ref{Energy-strong}, but for the weak scaling.}
    \label{Energy-weak}
    
\end{figure}

Figures~\ref{Runtime-weak} and~\ref{Energy-weak} show the weak-scaling results for runtime and energy consumption on the SETONIX cluster. All histograms are normalized and annotated as in Figure~\ref{Runtime-strong}. Runtime scaling is essentially flat, as expected for embarrassingly parallel codes where both the problem size and the number of {\CUs} increase together, with GPUs performing $\sim 8\times$ faster than CPUs, consistent with the strong-scaling results. 

Energy consumption, doubles at each step because the workload per \CU~is fixed while runtime remains flat: running the code for the same time while doubling resources implies a theoretical doubling of energy usage.

\begin{figure}[h]
    \centering
    \includegraphics[width=1\textwidth]{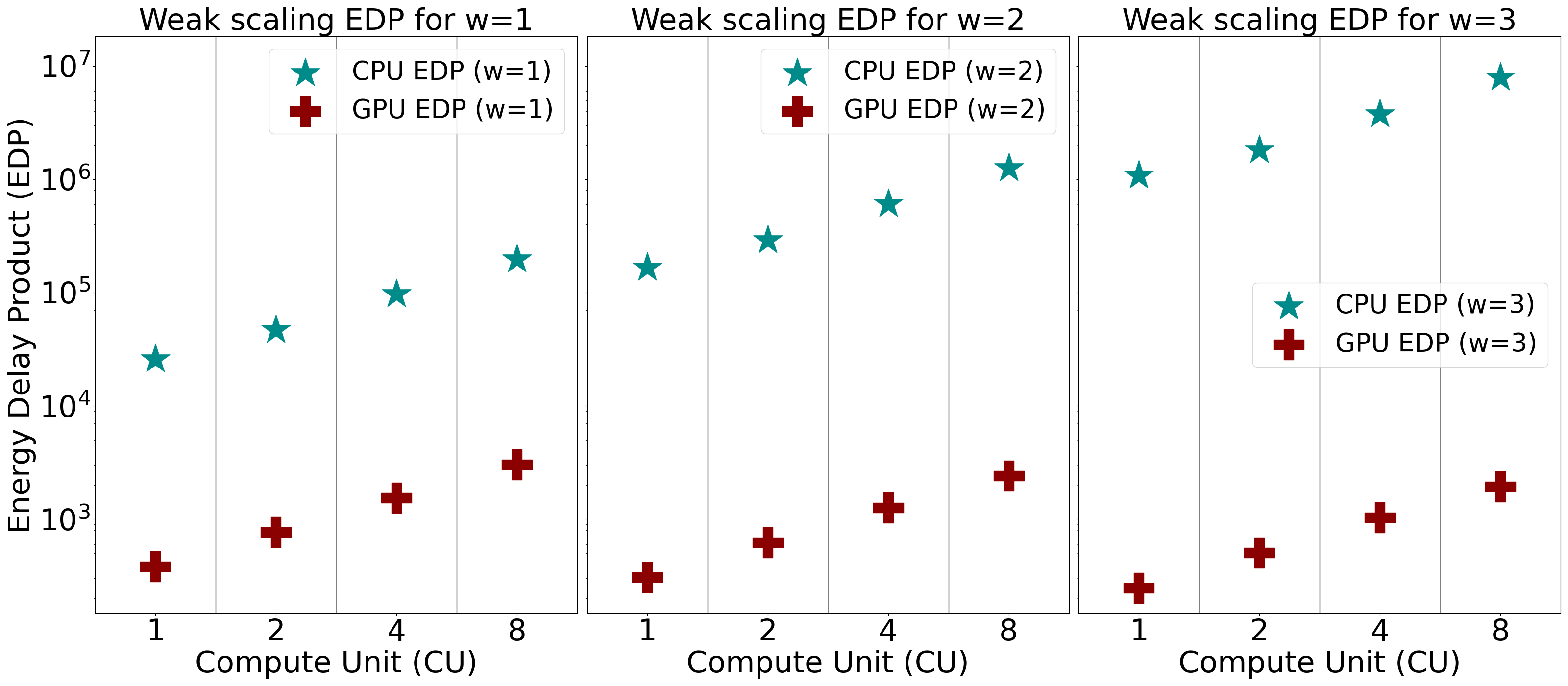}
    \caption{Same as Figure~\ref{EDP-strong}, but for weak scaling.}
    \label{EDP-weak}
    
\end{figure}

In Figure \ref{EDP-weak} we show the  results as in Figure~\ref{EDP-strong}, but for weak scaling. Unlike the trend we observed in Figure \ref{EDP-strong}, the \EDP~increases as a function of {\CUs} in this case. This is because the runtime remains constant while energy doubles at each step, leading to an overall increase of the \EDP~whenever the number of \CUs~is doubled. For $w=1$, GPUs are more efficient by a factor of $\sim 64\times$, consistent with the strong scaling results. For $w=2,3$, which correspond to cases where more weight is given to time-to-solution than energy-to-solution, the gap between GPUs and CPUs becomes increasingly pronounced.

\subsubsection{Green Productivity}
\label{GP-sub}
\GP~provides insights into how the increasing resources affects overall system utilization. Here,  we focus on the strong scaling tests discussed in Section~\ref{benchmarks} and summarized in Table~\ref{table:strong-scaling-table}, adopting $\alpha=1$ to give equal weight to energy-to-solution and time-to-solution. 

\begin{figure}[h]
    \centering
    \includegraphics[width=1\textwidth]{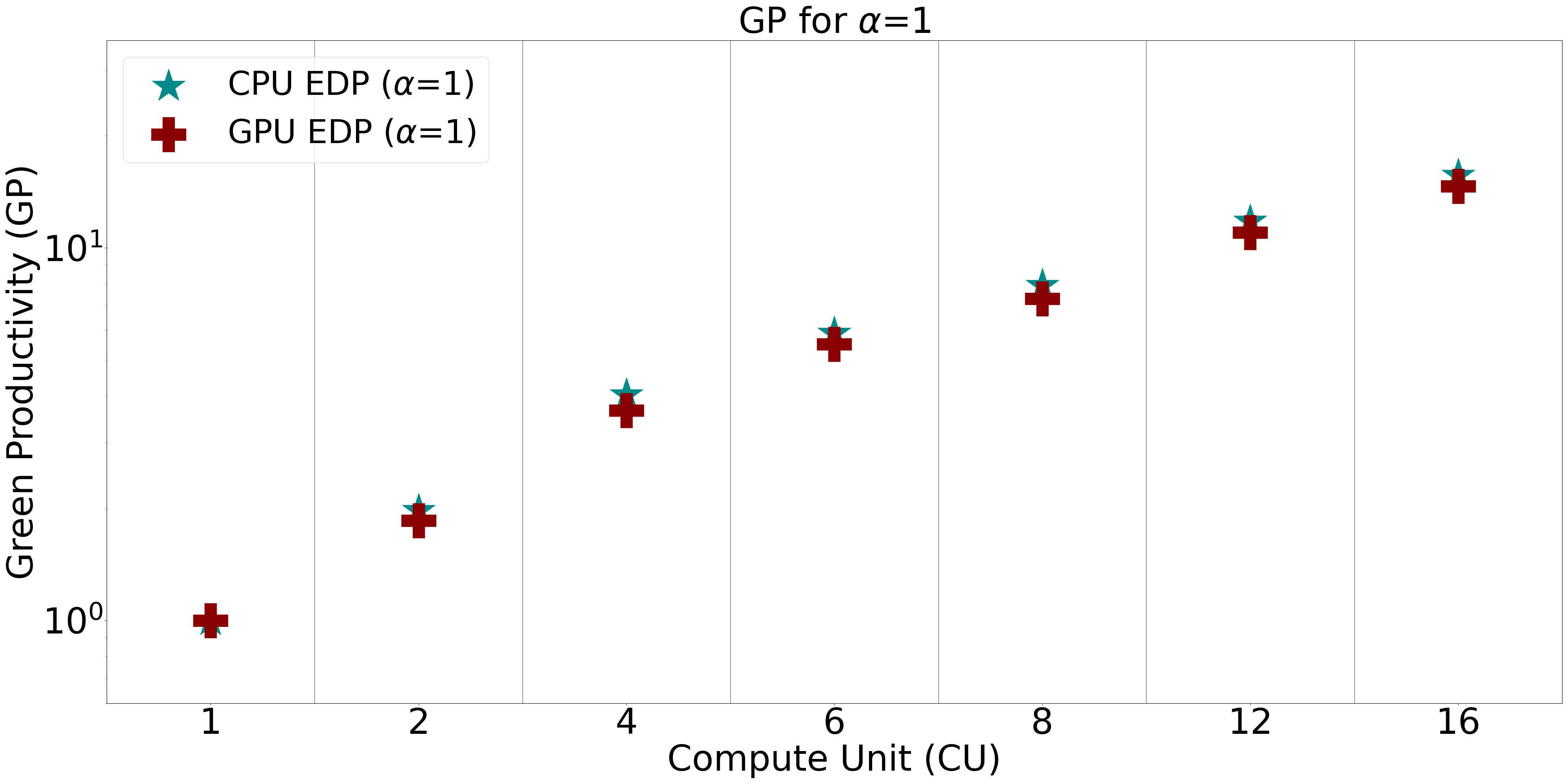}
    \caption{\GP~for the SETONIX cluster strong scaling tests, for all CPU and GPU configurations. For CPU runs, the reference is 1 CPU \CU, while for GPU ones the reference is 1 GPU \CU. Results are shown for $\alpha = 1$.}
    \label{GP-strong}
    
\end{figure}

Results for the SETONIX system are shown in Figure~\ref{GP-strong}, where cyan stars denotes CPU runs and dark red Greek crosses correspond to GPU runs. CPU and GPU runs are self-normalized: for CPU runs the reference configuration is the 1 CPU \CU, while for GPU runs the reference is 1 GPU \CU. Since the code is embarrassingly parallel, increasing the number \CUs~should in principle lead to higher \GP. However, this does not generally hold for non-compute bound algorithms, where the configuration that maximizes \GP~is often the one with the lowest \CUs~fitting the problem, as presented in \cite{lacopo2025greencomputingskaera}. 

Nevertheless, the results in Figure~\ref{GP-strong} shows that \GP~does not increase linearly, with a shallow ``knee"  appearing around 6 \CUs. This behavior indicates that the problem size becomes too small to fully utilize the available computing resources, especially for GPU runs.

\subsection{Karolina cluster}
\label{karolina-res}

\subsubsection{Strong scaling tests}
\label{strong-scaling-nvidia}
\begin{figure}[h]
    \centering
    \includegraphics[width=1.0\textwidth]{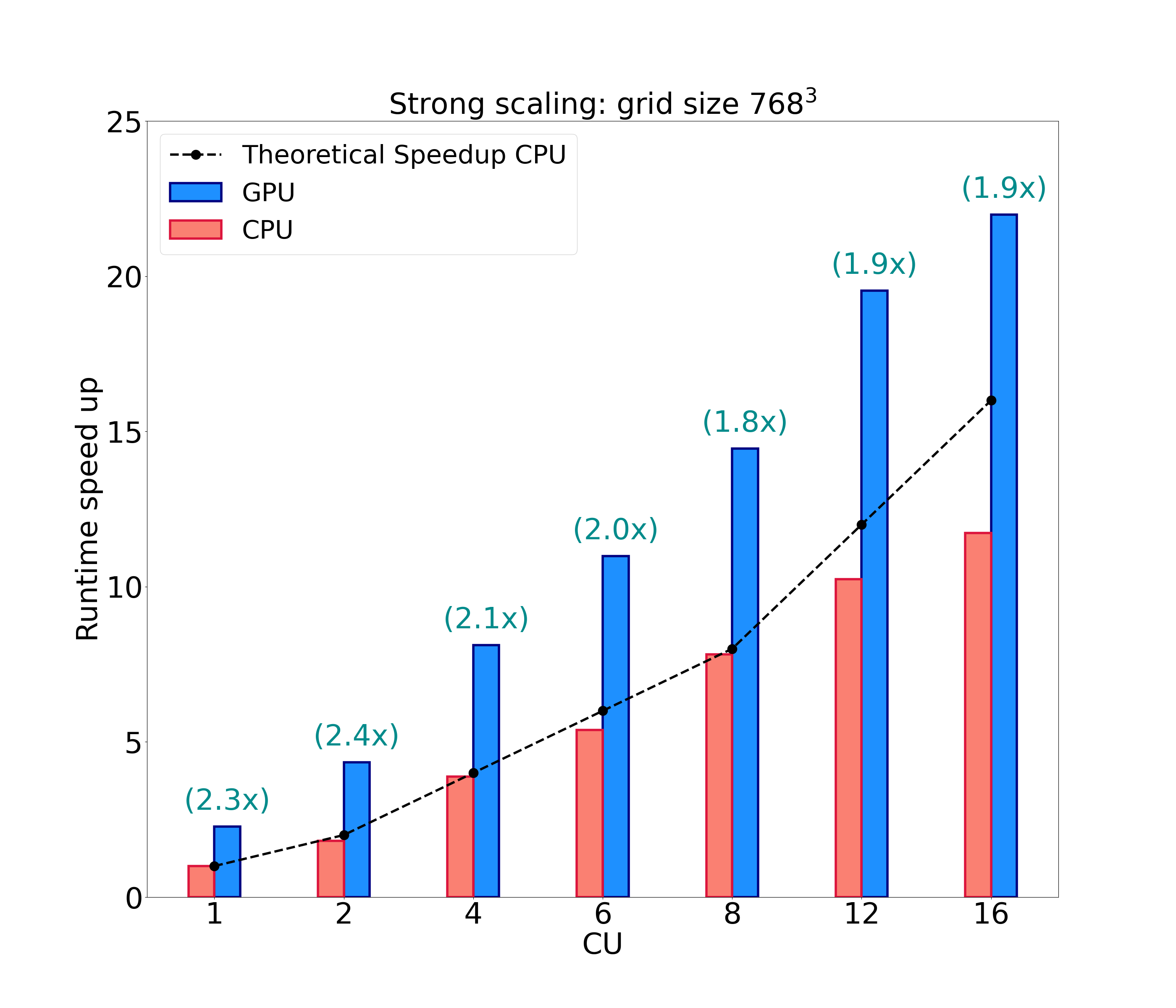}
    \caption{Same as Figure \ref{Runtime-strong}, but for the KAROLINA cluster.}
    \label{Runtime-strong-nvidia}
    
\end{figure}

Similar to Figure~\ref{Runtime-strong}, Figure~\ref{Runtime-strong-nvidia} presents the runtime strong-scaling results for KAROLINA. In this case, the runtime gain from GPU runs is about $4\times$ smaller than that observed on SETONIX.  This is due  both to higher performance of the dual-socket CPUs on KAROLINA (essentially twice the computing power of SETONIX CPUs) and to the difference in FP64 computing power between AMD and NVIDIA GPUs, with the latter offering lower peak performance (see Section~\ref{Computing_platforms}). As a result, the runtime gain for GPU configurations is limited to about $2\times$, compared to the homologous CPU configurations. Interestingly, KAROLINA exhibits worse strong scaling behavior than SETONIX for both CPU and GPU runs, with a significant deviation from the ideal scaling.

\begin{figure}[h]
    \centering
    \includegraphics[width=1.0\textwidth]{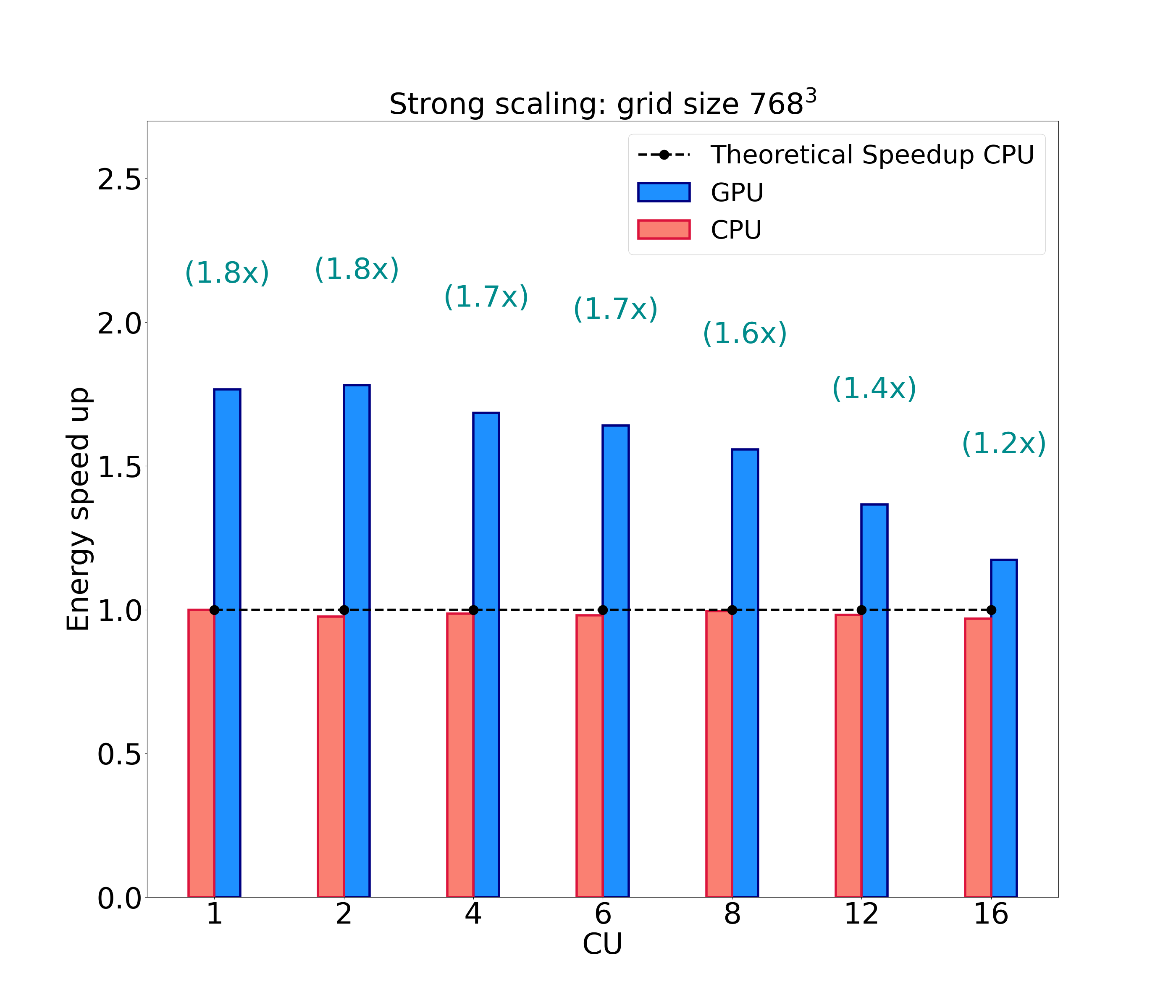}
    \caption{Same as Figure \ref{Energy-strong}, but for the KAROLINA cluster.}
    \label{Energy-strong-nvidia}
    
\end{figure}

However, GPU configurations show an even worse speedup than CPU configurations, as is particularly evident in the energy speedup result shown in Figure \ref{Energy-strong-nvidia}. For  runs with $12$ or $16$ {\CUs}, the energy gain factor approaches a $1\times$ factor, indicating that the GPU utilization becomes inefficient in these cases. Ideal strong scaling is achieved only as long as hardware is fully utilized. Beyond $12$ \CUs, this condition is no longer met for GPU tests. As a result,
GPUs remain underutilized while the runtime is dominated by CPU-GPU latency. The faster interconnection available on SETONIX partially mitigates for this under-utilization effect (see Section \ref{Computing_platforms}). 

\begin{figure}[h]
    \centering
    \includegraphics[width=1\textwidth]{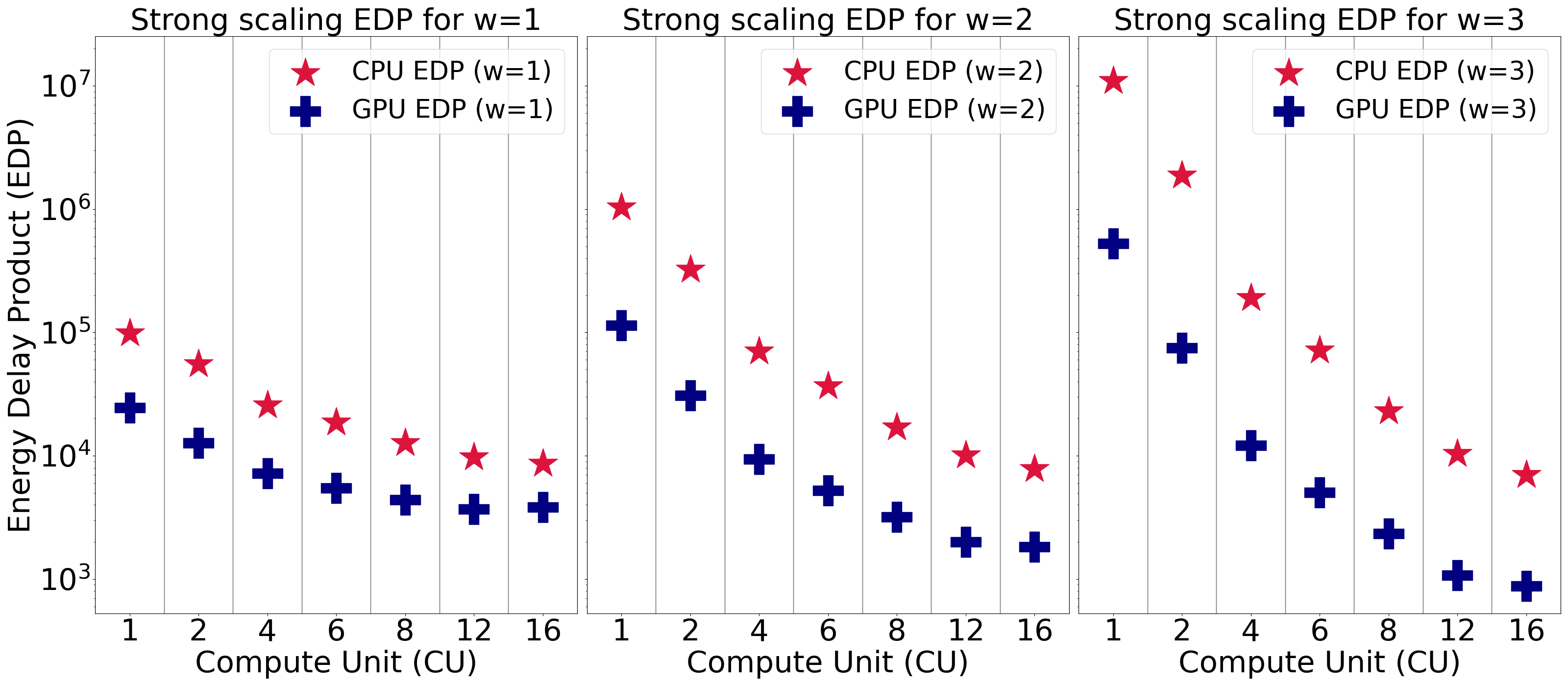}
    \caption{Same as Figure \ref{EDP-strong}, but for the KAROLINA cluster.}
    \label{EDP-strong-nvidia}
    
\end{figure}

\EDP~results for KAROLINA are shown in Figure~\ref{EDP-strong-nvidia}. Red stars refer to pure CPU runs, blue Greek crosses refer to GPU ones. In Figure~\ref{EDP-strong} we observed a $64\times$ advantage of GPUs over CPUs runs in the combined energy-runtime efficiency for $w=1$. ON KAROLINA, this gap shrinks to $4\times$ and decreases further for larger numbers of \CUs. As expected, the gap increases when $w=2,3$, but it remains less pronounced than in Figure \ref{EDP-strong}. 

\subsubsection{Weak scaling tests}
\label{weak-scaling-nvidia}
\begin{figure}[h]
    \centering
    \includegraphics[width=1.0\textwidth]{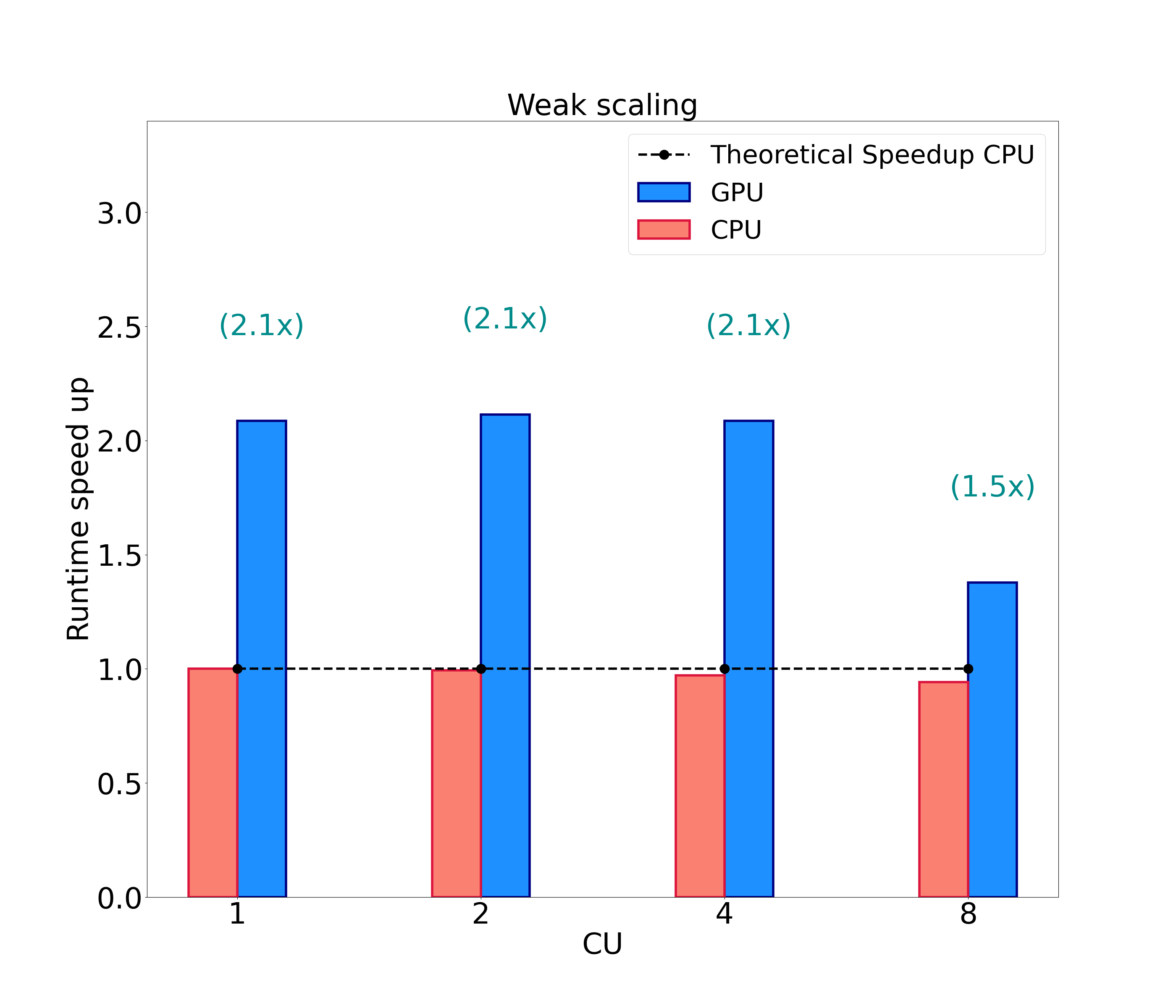}
    \caption{Same as Figure \ref{Runtime-weak}, but for the KAROLINA cluster.}
    \label{Runtime-weak-nvidia}
\end{figure}

Similar to Figure~\ref{Runtime-weak}, Figure~\ref{Runtime-weak-nvidia} presents the runtime weak-scaling results for KAROLINA. While CPU runtimes scale almost as expected, with only a slight deviation for 8 \CUs, the behavior with GPUs is more interesting. Specifically, when 8 \CUs~are used, the runtime gain factor drops from $2\times$ to $1.5\times$, following a trend similar to the strong scaling tests. This represents a significant deviation from theoretical scaling, indicating that GPUs become less efficient in terms of runtime when many computing nodes are used. 

\begin{figure}[h]
    \centering
    \includegraphics[width=1.0\textwidth]{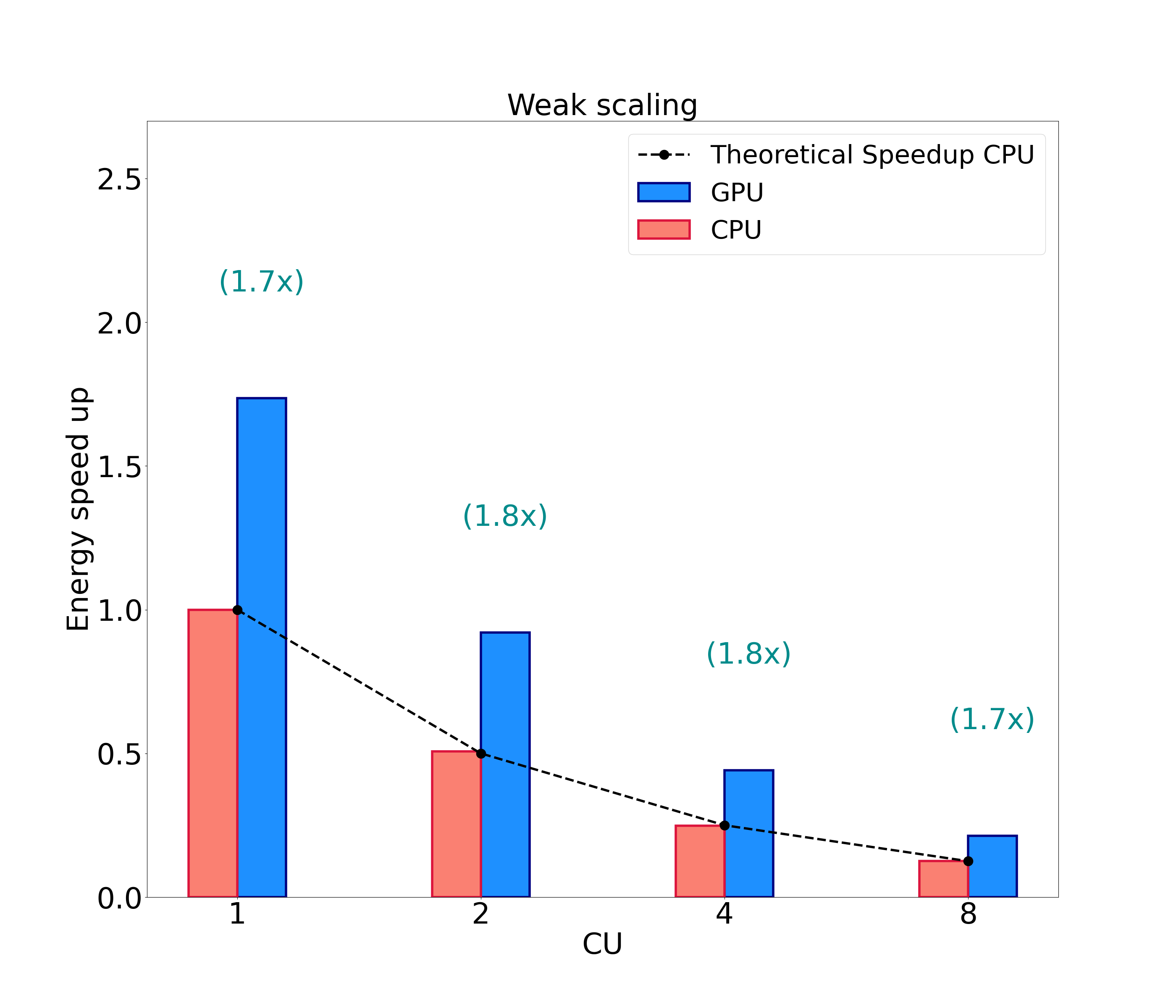}
    \caption{Same as Figure \ref{Energy-weak}, but for the KAROLINA cluster.}
    \label{Energy-weak-nvidia}
    
\end{figure}

Similar to Figure~\ref{Energy-weak}, Figure~\ref{Energy-weak-nvidia} presents the energy weak-scaling results for KAROLINA. Interestingly, the energy gain factors of each GPU configurations relative to the corresponding CPU configurations are stable, without the significant drop observed in Figure ~\ref{Runtime-weak-nvidia}. This suggests that the energy is not simply an integral over time. On some system, faster runtimes do not necessarily mean greener execution, and vice-versa. Overall, weak scaling tends to behave as expected in both CPU and GPU runs.

\begin{figure}[h]
    \centering
    \includegraphics[width=1\textwidth]{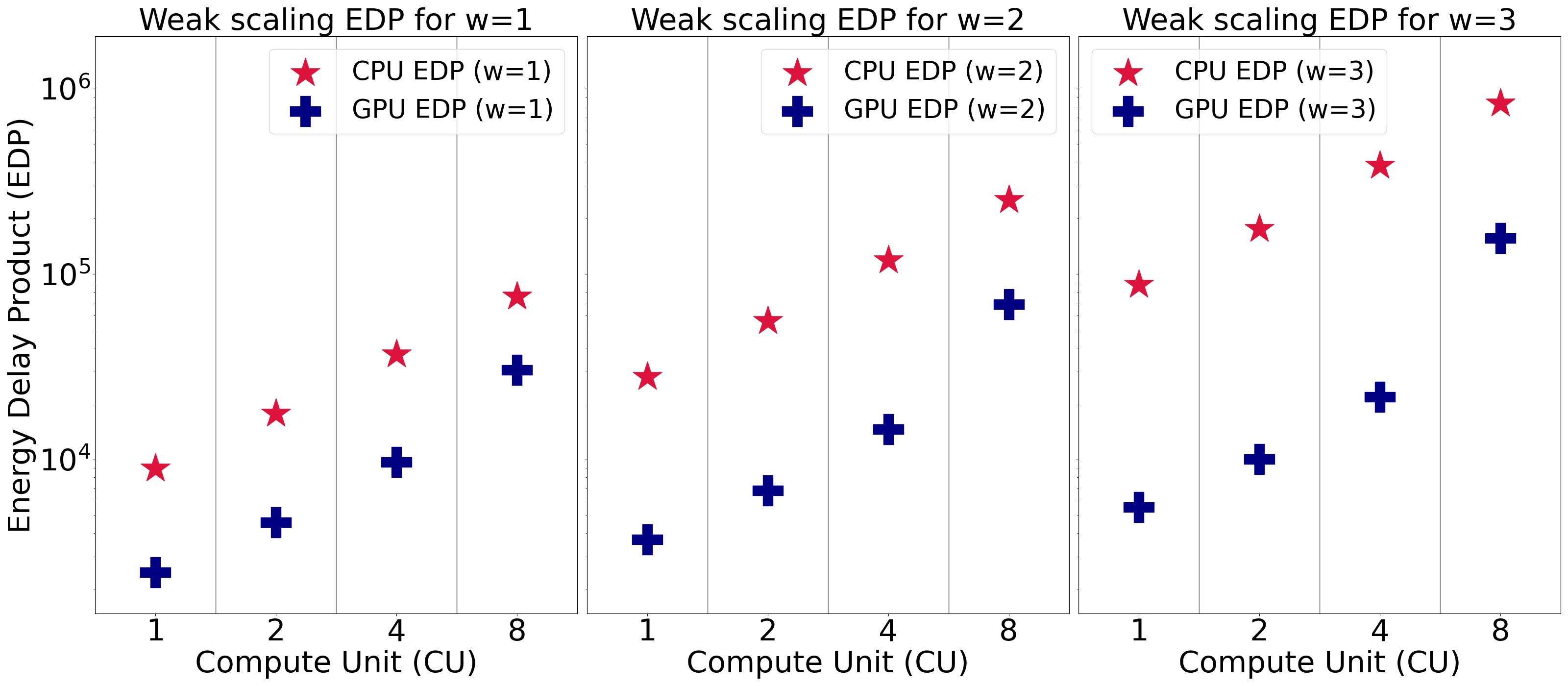}
    \caption{Same as Figure \ref{EDP-weak}, but for the KAROLINA cluster.}
    \label{EDP-weak-nvidia}
    
\end{figure}

Weak scaling \EDP~ results for KAROLINA are shown in Figure \ref{EDP-weak-nvidia}. The drop in GPU efficiency from $64\times$ to $4\times$ is consistent with strong scaling results in Figure \ref{EDP-strong-nvidia} for $w=1$, with a drop at 8 \CUs. As expected, the gap between GPUs and CPUs widens for $w=2,3$. 

\subsubsection{Green Productivity}
\label{GP-sub-nvidia}

\begin{figure}[h]
    \centering
    \includegraphics[width=1\textwidth]{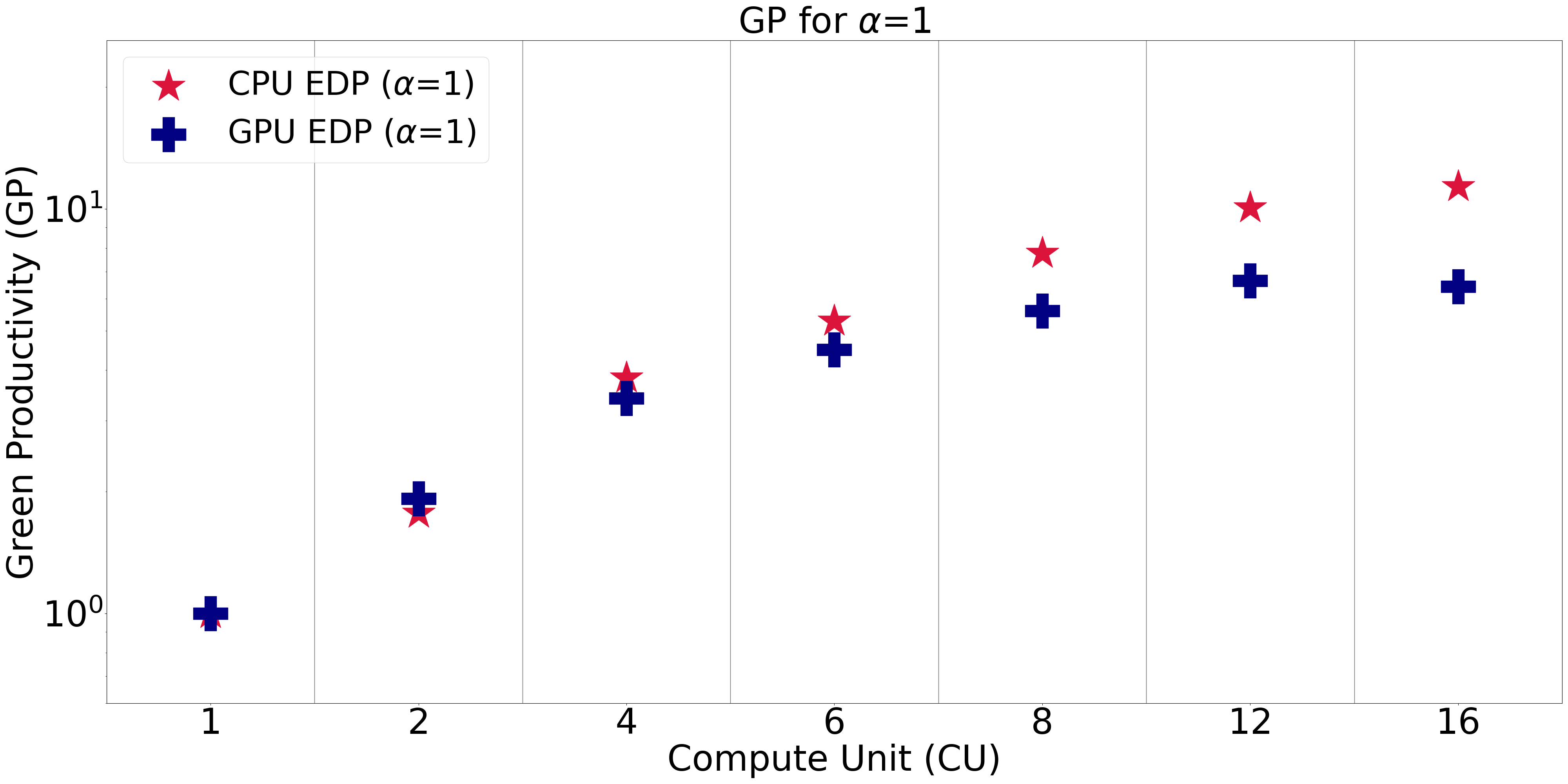}
    \caption{Same as Figure \ref{GP-strong}, but for the KAROLINA cluster.}
    \label{GP-strong-nvidia}
    
\end{figure}

\GP\ results for the KAROLINA cluster, shown in Figure~\ref{GP-strong-nvidia}, were obtained following the same approach described in Section~\ref{GP-sub}, but using the configurations listed in Table~\ref{table:strong-scaling-table-nvidia}. Unlike in Figure~\ref{GP-strong}, where \GP~steadily increasing with \CUs, on KAROLINA the non-ideal scaling results in a plateau as the number of computing resources increases. This effect is even more pronounced in GPU runs, where a drop is observed for runs with more than than 12 \CUs. 

\subsection{Karolina vs Setonix \EDP}
\label{sp:norm_edp}
\begin{figure}[h]
    \centering
    \includegraphics[width=1\textwidth]{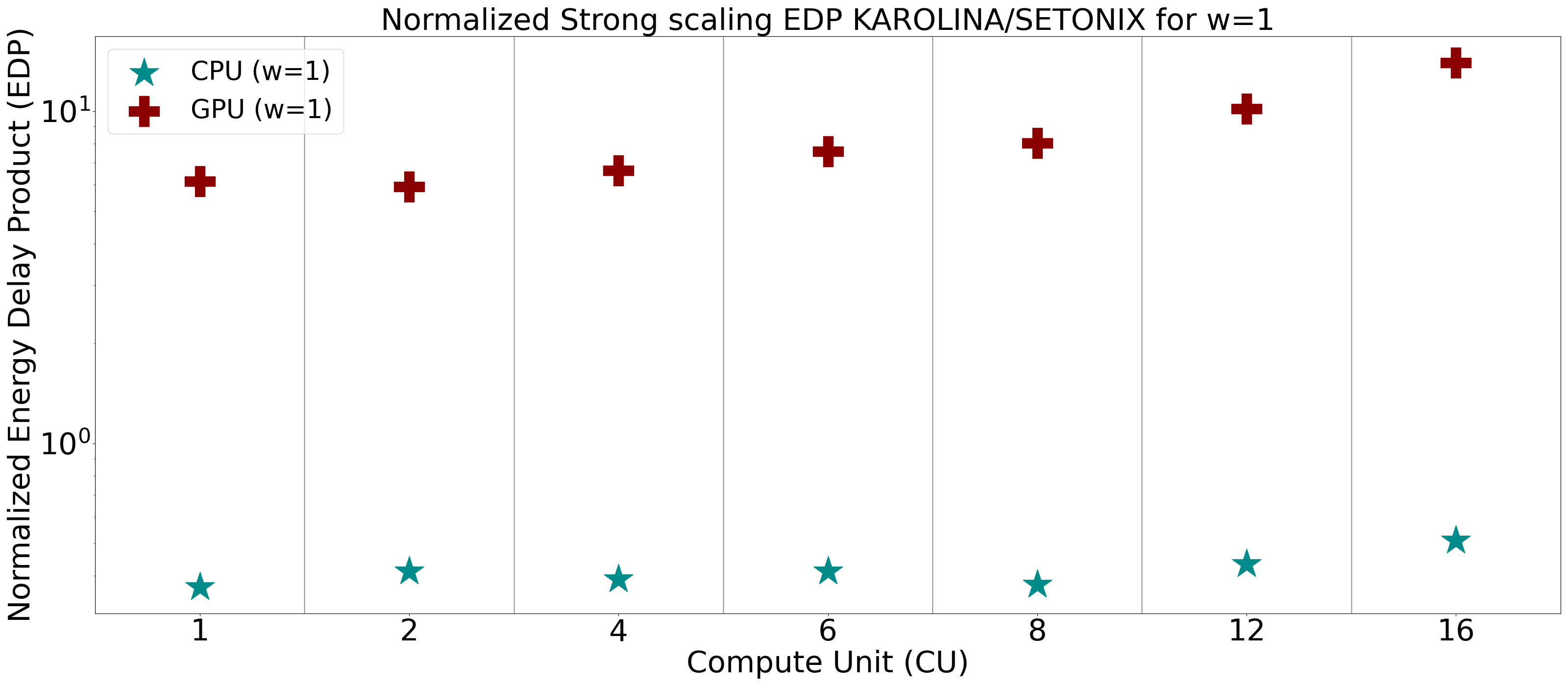}
    \caption{Normalized strong scaling \EDP~ of the two clusters, obtained with the ratio between KAROLINA \EDP~and SETONIX \EDP, for CPU and GPU configurations.}
    \label{fig:norm_strong}
    
\end{figure}

\begin{figure}[h]
    \centering
    \includegraphics[width=1\textwidth]{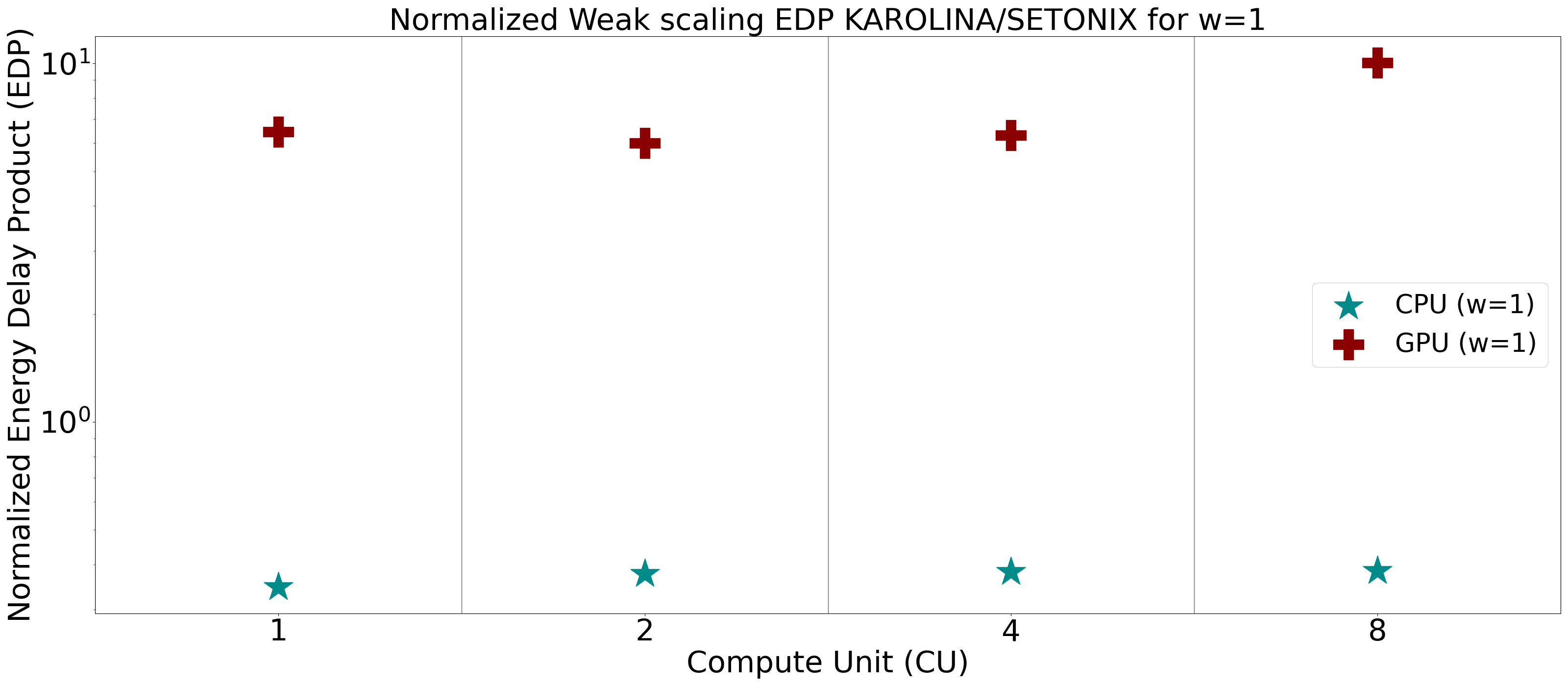}
    \caption{Same as Figure \ref{fig:norm_strong}, but for the normalized weak scaling \EDP.}
    \label{fig:norm_weak}
    
\end{figure}

To infer which out of the two architectures is best suited for the \kernel\ under investigation, it is useful to analyzed the normalized \EDP, i.e. the ratio between KAROLINA and SETONIX \EDPs. Figure~\ref{fig:norm_strong} shows the normalized \EDP, with $w=1$, for strong scaling tests. The CPU \EDP~ratio is lower than one, meaning that KAROLINA is more efficient for CPU-only runs. This is expected, since KAROLINA has twice CPU cores than SETONIX, whereas for the GPU configurations SETONIX is more efficient by an order of magnitude. This is mainly a result of a better FP64 performance available in AMD GPUs compared to NVIDIA. For the sake of completeness, we show the normalized weak scaling \EDP, with $w=1$, in Figure~\ref{fig:norm_weak}. Similar to the strong scaling case, KAROLINA best fits for pure CPU applications, while SETONIX is much better for GPU ones. However, we stress out that this specific analysis is not portable, and such plots should be provided for each couple of HPC platforms in which each user is supposed to run a code.

\section{Conclusions}
\label{sec:conclusions}
In this paper, we assessed the energy efficiency of the GPU-accelerated version of {\pinocchio}, a widely used tool for generating mock halo catalogues, with a dual focus on maximizing performance and minimizing energy consumption. To this end, we measured energy-to-solution alongside time-to-solution using a new parallel implementation of the PMT (Section~\ref{parallel-pmt}) across different HPC platforms equipped with AMD and NVIDIA GPUs (Section~\ref{Computing_platforms}). Our primary goal was to address the growing need for computational sustainability in astrophysics, particularly in the context of producing large ensembles of mock halo catalogues for upcoming surveys such as \textit{Euclid}.

On the AMD-based system, benchmarks show an $\sim 8\times$ improvement in both time-to-solution and energy-to-solution relative to the CPU-only implementation (Sections~\ref{strong-scaling} and~\ref{weak-scaling}), yielding an overall efficiency gain of $\sim 64\times$, as reflected in a substantially lower EDP (Figures~\ref{EDP-strong} and~\ref{EDP-weak}). On the NVIDIA-based system, the improvements are more modest, with $\sim 2\times$ gains in both runtime and energy consumption, corresponding to a net efficiency increase of $\sim 4\times$ (Sections~\ref{strong-scaling-nvidia} and~\ref{weak-scaling-nvidia}).

The primary performance difference is mainly due to the disparity in peak double-precision (FP64) throughput between the AMD and NVIDIA architectures. Specifically, the AMD GPU exhibits $47.9 \text{ TFlops}$ peak FP64 performance ($23.9 \text{ TFlops}$ per Graphics Compute Die (GCD)), significantly contrasting with the $9.7 \text{ TFlops}$ demonstrated by the NVIDIA counterparts. This gap is critically amplified by the unavoidable necessity of double-precision arithmetic in contemporary cosmological simulations, which leverages the $1:1$ single-to-double precision (FP32:FP64) ratio of the AMD accelerators, versus the $2:1$ ratio of the NVIDIA accelerators. The central objective of this study is to demonstrate cross-platform code portability across both device ecosystems. It is crucial to note, however, that code portability does not inherently ensure performance portability across different architectures, with the superior AMD performance constituting a key achievement of this investigation. Nonetheless, any exhaustive performance analysis must rigorously account for the system-level architectural heterogeneity, including factors such as GPU/CPU count and the specific CPU-GPU interconnect topology, which dramatically impact overall efficiency and change the relative efficiency between one cluster and another one. The analysis of normalized \EDP~reveals which machine is best to run our \kernel~in CPU-only configurations and in GPUs, and we inferred that, depending on the configuration, one platform might be better than the other one and vice versa. Determining which platform is more efficient for a specific code is a non trivial task, and even for the same couple of architectures, it may differ depending on the specific configuration set up.

 Taken together, these results confirm that the GPU-accelerated version of {\pinocchio} is not only significantly faster than its CPU-based predecessor but also demonstrably more energy-efficient. 

Overall, GPU acceleration consistently reduces runtime and energy consumption, although the magnitude of the gains is system-dependent (reinforcing the need for portability), thereby enhancing the sustainability of large-scale cosmological campaigns.


This achievement showcases a viable and effective pathway for modernizing legacy scientific codes to meet the demands of the exascale era. By leveraging the parallel processing power of GPUs, we enhance the overall green productivity of cosmological simulations, enabling the generation of vast datasets required for next-generation surveys with a minimized environmental impact. This allows us to tackle larger and more complex cosmological problems without a proportional increase in our energy footprint. The methodology and results presented here serve as a compelling case study for the broader scientific community, highlighting the importance of adopting green-computing practices.

Determining the optimal execution environment for a given computational workload, particularly regarding the efficiency of GPU acceleration, is a non-trivial challenge. In this study, all performance benchmarks were conducted on GPU-equipped nodes, thereby incorporating the idle energy consumption of the accelerators into our analysis. This approach reflects a realistic scenario where GPUs, as integral components of the node, draw power regardless of their operational state. However, many supercomputing centers offer CPU-only partitions, presenting an alternative for workloads that may not benefit from GPU offloading. We propose that a systematic analysis of metrics such as the \EDP~and \GP~can serve as a valuable decision-making tool for HPC facility managers. By visualizing the trade-offs between performance and energy consumption, these metrics can guide the allocation of computational resources, especially in cases where the \EDP~values for CPU and GPU executions are comparable and the \GP~fails to exhibit strong scaling. Adopting such a quantitative approach to resource allocation can significantly mitigate energy waste, promoting a more sustainable and cost-effective operation of HPC infrastructure.

Future work will involve extending these optimizations to other components of our simulation pipelines and exploring further algorithmic improvements to push the boundaries of sustainable computational cosmology.

\section*{Acknowledgment}
This work has been supported by the Spoke-1 and Spoke-3 ``FutureHPC \& BigData” of the ICSC – Centro Nazionale di Ricerca in High Performance Computing, Big Data and Quantum Computing – and hosting entity, funded by European Union – NextGenerationEU. Supported by Italian Research Center on High Performance Computing Big Data and Quantum Computing (ICSC), project funded by European Union - NextGenerationEU - and National Recovery and Resilience Plan (NRRP) - Mission 4 Component 2 within the activities of Spoke 2 (Fundamental Research and Space Economy), (CN 00000013 - CUP C53C22000350006). We acknowledge the Pawsey Supercomputing Research Centre and the IT4Innovations National Supercomputing Center for the availability of high performance computing resources, support and collaborations. This work has also been supported by the National Recovery and Resilience Plan (NRRP), Mission 4, Component 2, Investment 1.1, Call for tender No. 1409 published on 14.9.2022 by the Italian Ministry of University and Research (MUR), funded by the European Union – NextGenerationEU– Project Title ``Space-based cosmology with Euclid: the role of High-Performance Computing” – CUP J53D23019100001 - Grant Assignment Decree No. 962 adopted on 30/06/2023 by the Italian Ministry of Ministry of University and Research (MUR).

\bibliographystyle{elsarticle-harv} 
\bibliography{biblio}

\appendix
\section{Technical specifications of the two architectures}
\label{architecture_appendix}
This Appendix provides a detailed description of the two HPC platforms selected for this study, KAROLINA and SETONIX.

\subsection{CPU-GPU Interconnection Topology}
\begin{figure}[h]
    \centering
    \includegraphics[width=1\textwidth]{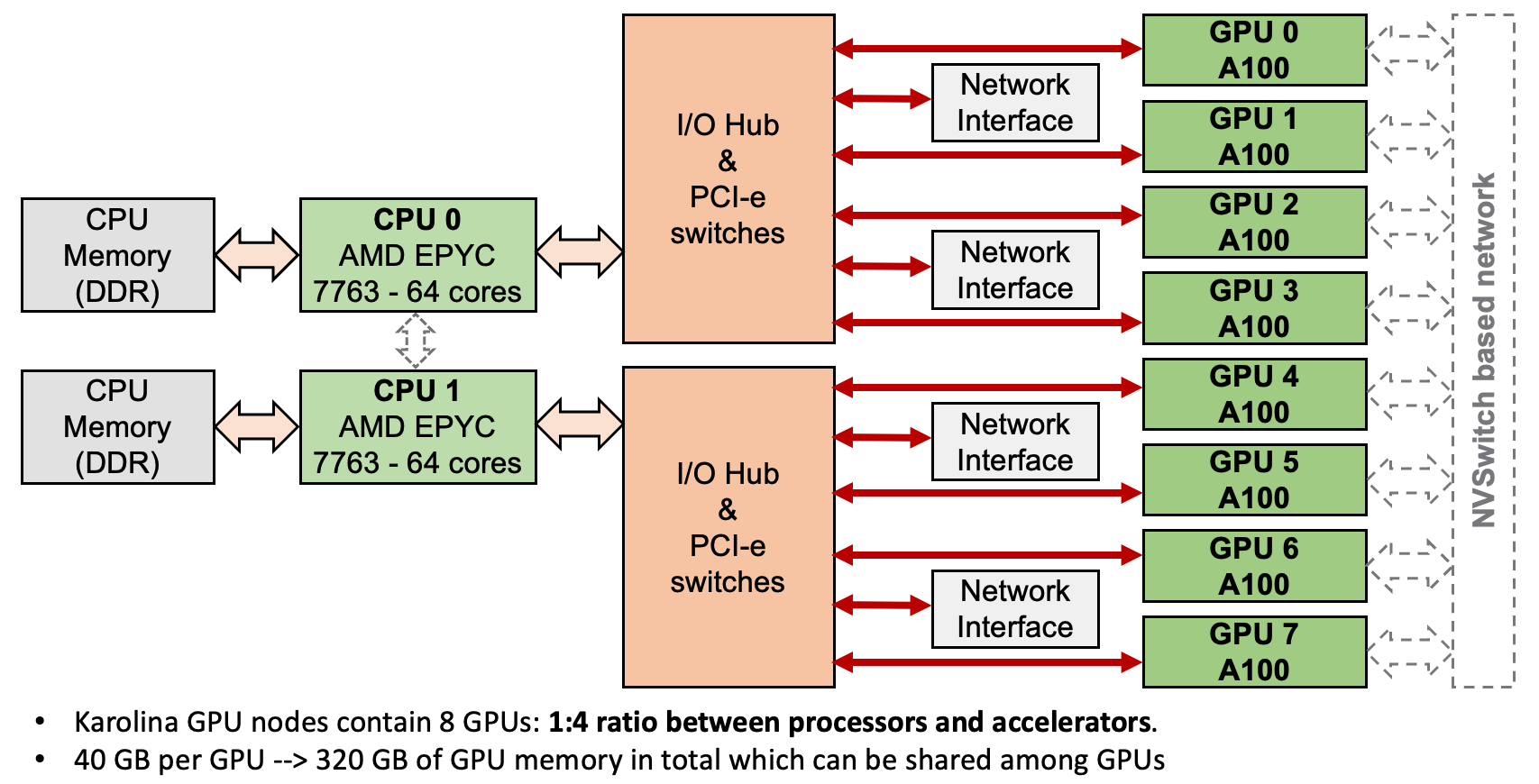}
    \caption{KAROLINA-GPU cluster architecture. GPUs are connected to sockets, constituted by 64 CPU cores each. There are 8 GPUs and two sockets, meaning that 4 GPUs are connected to each socket. GPUs are connected to the corresponding sockets through PCIe 4.0 (31.5 GB/s per link). Each pair of GPUsis connected through NVSwitch at bidirectional bandwidth of 600 GB/s. Credits: Lubomir Riha.}
    \label{Karolina-structure}
\end{figure}

The KAROLINA compute nodes adopt a conventional PCIe-based interconnect architecture. A schematic representation of one compute node is shown in Figure \ref{Karolina-structure}. The node specifications are described in detail below:

\begin{equation*}
\begin{split}
\text{CPU Socket 0 (64 cores)} &\xrightarrow{\text{PCIe Gen4}} \text{GPU 0, GPU 1, GPU 2, GPU 3} \\
\text{CPU Socket 1 (64 cores)} &\xrightarrow{\text{PCIe Gen4}} \text{GPU 4, GPU 5, GPU 6, GPU 7}
\end{split}
\end{equation*}
\begin{itemize}
    \item \textbf{Host-to-Device Interconnect}: The primary communication channel employs PCI Express Generation 4 ($\text{PCIe Gen4}$) $\times 16$ links, delivering a $31.5 \text{ GB/s}$ bidirectional bandwidth per GPU. The associated data transfer latency is nominal, typically ranging from $1$ to $2 \text{ \textmu s}$.
    \item \textbf{Non-Uniform Memory Access (NUMA) Considerations}: The dedicated attachment of accelerator groups to specific CPU sockets required an optimized Message Passing Interface (MPI) rank placement. This strategy is crucial for mitigating cross-socket traffic to the host memory, thereby preserving memory access performance.
    \item \textbf{Device-to-Device Interconnect}: While the system is equipped with NVSwitch, capable of providing a $600 \text{ GB/s}$ inter-GPU bandwidth between paired devices, this high-speed link remains unused in the context of the current embarrassingly parallel workload.
\end{itemize}

\begin{figure}[h]
    \centering
    \includegraphics[width=1\textwidth]{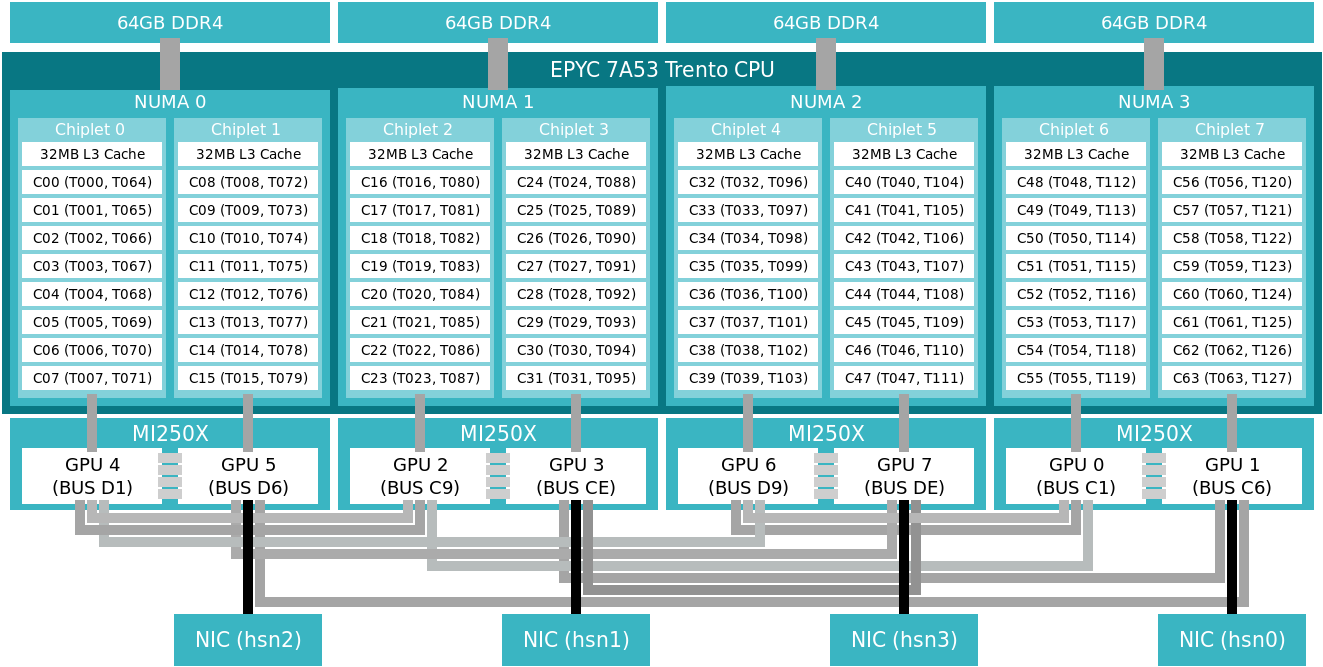}
    \caption{Setonix-GPU cluster architecture. Each GCD is assigned to a particular chiplet, constituted by 8 CPU cores each. GPUs are connected to the corresponding chiplets through InfinityFabric (36-50GB per link). GCDs in the same GPU are connected through InfinityFabric at bidirectional bandwidth of 400 GB/s, while GCDs in different GPUs have 200 GB/s bidirectional bandwidth connections. It should be noted that chiplet IDs do not, in general, correspond directly to the IDs of the GCDs. Credits: \url{https://pawsey.atlassian.net/wiki/spaces/US/pages/51928618/Setonix+GPU+Partition+Quick+Start}.}
    \label{Setonix-structure}
\end{figure}

The SETONIX compute nodes are architecturally based on the AMD unified Infinity Fabric (IF) architecture, which governs the intricate interconnectivity. The resulting NUMA structure is conceptually represented in Figure \ref{Setonix-structure}.

Each AMD Optimized 3rd Generation EPYC ``Trento" CPU (comprising 64 total cores) is logically segmented into eight chiplets (eight cores per chiplet). The specific coupling between CPU chiplets and their respective GCDs is mediated via the IF, as mapped below:

\begin{equation*}
\begin{split}
\text{Chiplet 0 (8 cores)} &\xrightarrow{\text{Infinity Fabric}} \text{GCD 4} \\
\text{Chiplet 1 (8 cores)} &\xrightarrow{\text{Infinity Fabric}} \text{GCD 5} \\
\text{Chiplet 2 (8 cores)} &\xrightarrow{\text{Infinity Fabric}} \text{GCD 2} \\
\text{Chiplet 3 (8 cores)} &\xrightarrow{\text{Infinity Fabric}} \text{GCD 3} \\
\text{Chiplet 4 (8 cores)} &\xrightarrow{\text{Infinity Fabric}} \text{GCD 6} \\
\text{Chiplet 5 (8 cores)} &\xrightarrow{\text{Infinity Fabric}} \text{GCD 7} \\
\text{Chiplet 6 (8 cores)} &\xrightarrow{\text{Infinity Fabric}} \text{GCD 0} \\
\text{Chiplet 7 (8 cores)} &\xrightarrow{\text{Infinity Fabric}} \text{GCD 1}
\end{split}
\end{equation*}
\begin{itemize}

    \item \textbf{Bandwidth Specification}: The Infinity Fabric link provides a high-throughput channel with a bandwidth ranging from $36$ to $50 \text{ GB/s}$ per link, subject to the specific configuration utilized.
    \item \textbf{Data Transfer Latency}: Due to the tightly integrated system design, the interconnect achieves sub-microsecond latency for data transfers.
    \item \textbf{Memory Coherence}: The system implements a unified memory architecture, which inherently supports coherent memory access between the CPU's host memory space and the GPU's device memory space.
    \item \textbf{Topology and Locality}: This architecture establishes a natural affinity where each 8-core CPU chiplet maintains a direct, dedicated connection to its paired GCD. This tight integration eliminates conventional cross-socket NUMA access concerns for compute tasks utilizing the local CPU-GCD pair.
\end{itemize}
\subsection{Performance Implications for Our Workload}
The disparities between the two computing architectures introduce several critical factors impacting the performance and efficiency of the \textit{kernel} workload:

\begin{enumerate}
    \item \textbf{Arithmetic Intensity and Throughput}: The AMD MI250X accelerator provides a significantly higher peak $\text{FP64}$ throughput ($47.9 \text{ TFLOPS}$ versus $9.7 \text{ TFLOPS}$). For the present compute-bound workload, this disparity directly translates to an accelerated kernel execution time.
    
    \item \textbf{Memory Subsystem Throughput}: While the AMD platform offers substantially greater High Bandwidth Memory (HBM) aggregate throughput ($3,276 \text{ GB/s}$ versus $1,555 \text{ GB/s}$), this performance differential is less consequential for the primary compute-bound phase of the kernel.
    
    \item \textbf{Host-Device Data Transfer Overheads}: The interconnect technology dictates the latency and bandwidth characteristics for host-to-device communication:
    \begin{itemize}
        \item \textbf{KAROLINA}: The $\text{PCI Express Generation 4}$ ($\text{PCIe Gen4}$) interface constrains peak transfer rates and incurs higher communication latency.
        \item \textbf{SETONIX}: The tightly integrated $\text{Infinity Fabric}$ architecture effectively minimizes transfer overhead, which is particularly advantageous for workloads characterized by frequent, low-volume data exchanges.
    \end{itemize}
    
    \item \textbf{Scalability and Multi-Accelerator Performance}: The intra-node interconnect topology affects performance consistency under full utilization:
    \begin{itemize}
        \item \textbf{KAROLINA}: The $8 \text{ GPU}$ per node configuration, relying on a $\text{PCIe}$ root complex, is prone to interconnect congestion and performance degradation when all accelerators attempt concurrent data movement operations.
        \item \textbf{SETONIX}: The $8 \text{ GCD}$ configuration, leveraging dedicated $\text{CPU-GPU}$ $\text{Infinity Fabric}$ links, facilitates more consistent performance scaling under maximum node load.
    \end{itemize}
    
    \item \textbf{Toolchain Maturity and Optimization Efficacy}: The state of the compiler ecosystem introduces a developmental factor:
    \begin{itemize}
        \item \textbf{NVIDIA}: Benefits from a highly mature compiler toolchain ($\text{NVC++ 24.3}$), offering established, extensive optimization for scientific $\text{HPC}$ workloads.
        \item \textbf{AMD}: Employs a newer toolchain ($\text{amdclang-18}$). Despite the last improvements, its overall optimization maturity and stability may not yet be equivalent to the $\text{NVIDIA}$ standard for all compiling scenarios.
    \end{itemize}
\end{enumerate}

\section{Parallel PMT technical details}
\label{pmt_appendix}
This Appendix focuses on the detailed description of how to use the Parallel PMT library in scientific applications.
PMT is compiled as a standard shared object (.so) that can be linked to any C/C++ code. To use it, the header must be included:

\begin{verbatim}
    #include "pmt.h"
\end{verbatim}

and compilation requires specifying the appropriate PMT \textit{include} and \textit{library} paths as follows:

\begin{verbatim}
    CC/CXX -I$(PATH_TO_PMT)/include -L$(PATH_TO_PMT)/lib64 
    my_code.c/my_code.cpp -lpmt
\end{verbatim}

Since PMT does not have a native MPI support, our effort has been put in the development of a MPI library utilizing PMT as a wrapper. Also the parallel library is compiled as a standard object.

Once the parallel library is compiled, the file:
\begin{verbatim}
    libpmt_parallel.so
\end{verbatim}
is created, and the header:
\begin{verbatim}
    #include "energy_pmt.h"
\end{verbatim}
must be included in the application. 

When the application starts, PMT is initialized through the function call:
\begin{verbatim}
    PMT_CREATE(MPI Comm comm, int rapl, int *devID, int numGPUs);
\end{verbatim}
where \texttt{comm} is the MPI communicator of the original application, \texttt{rapl} is an integer that can be either 0 (\texttt{rapl}  reading not active) or 1 (\texttt{rapl} reading active), \texttt{devID} is a pointer to the array of devices (per task) and \texttt{numGPUs} is the number of GPUs associated to each MPI task. In this work, we set \texttt{numGPUs} equal to 1, so that each MPI task is associated with a single GPU. In a future release of the library, we plan to extend the support to non trivial cases to where more than one GPU is exclusively assigned to a taks. 

The \texttt{PMT\_CREATE} function initializes the PMT counters for the specific hardware and the MPI communicators required by the library. Indeed, the library is MPI- and NUMA-aware: it automatically detects the number of computing nodes and to which node the specific task belongs. 

For multi-socket nodes, the library was extended to create communicators at the socket level, collecting all the MPI tasks running on the same socket. In addition to the single sockets and single nodes communicator, each socket and each node elects a master process (the rank 0 in the specific communicator). Higher-level communicators are then created to gather these socket and node masters. This structure enables meaningful statistical analyses, for example when running an application on many nodes and the goal is of computing averages and standard deviations across them. In the case of multi-socket nodes, statistics are available at both node and socket levels.

We discussed the possibility of having either single GPUs devices, such as the NVIDIA Tesla A100 (KAROLINA), or multi-GCD GPUs, such as the AMD Radeon Instinct MI250X (SETONIX). In the former case, the setup is straightforward: one MPI task is assigned to each GPU, and the task is responsible for accessing the hardware counters and exchanging information with all the other processes. The latter case is more complex, since the original application assigns one MPI task per GCD, but only one task per physical GPU should access the hardware counters. On AMD GPUs, this translates into two MPI ranks per GPU (one per GCD), with only the first rank reading the GPU counters and sharing the results with the other ranks. Consequently, in this specific case, only even MPI ranks initialize the PMT, while a dedicated communicator is initialized in the original code and passed to the library as \texttt{comm} argument.

Once PMT is initialized, the code segment or the \kernel~to be profiled is selected and bracketed as shown in the following pseudo-code:

\begin{verbatim}
    PMT_CPU_START("Kernel_name");
    PMT_GPU_START("Kernel_name", int devID);

    kernel execution

    PMT_CPU_STOP("Kernel_name");
    PMT_GPU_STOP("Kernel_name", int devID);
\end{verbatim}
where \texttt{devID} is the device number assigned to the specific MPI task. The tag must be the same for the \texttt{PMT\_*\_START} and \texttt{PMT\_*\_STOP} functions, otherwise PMT will return an error.

At the end of the run the global master process (the master of all PMT communicators), generates the results using  the following functions:
    
\begin{verbatim}
    PMT_CPU_SHOW("Kernel_name");
    PMT_GPU_SHOW("Kernel_name", int devID);
\end{verbatim}

Once this is done, the library automatically writes two output files per tag, one for CPUs and one for GPUs, reporting the total energy consumed by the entire \textit{kernel} as well as the statistics at the socket/node level. For GPUs, additional statistics are provided at the individual GPU level. 

\begin{figure}[h!]
    \centering
    \includegraphics[width=0.69\textwidth]{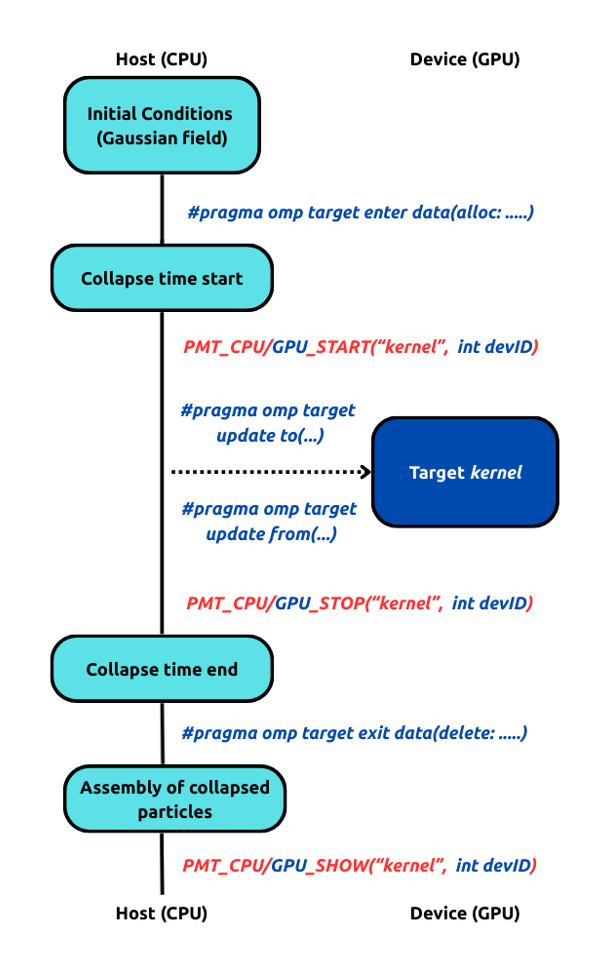}
    \caption{Schematic overview of the GPU porting of the {\pinocchio} collapse time \textit{kernel} and the integration of energy measurements. The diagram highlights the PMT function calls used for profiling, together with the data management and execution flow on the host and device.
    }
    \label{Energy_strategy}
\end{figure}

Finally, the following function must be called to free the memory allocated from the MPI communicators:
\begin{verbatim}
    PMT_FREE();
\end{verbatim}

For completeness, here is an example of how to compile and link a code with the parallel PMT:

Figure \ref{Energy_strategy} shows how to bracket the relevant code portions (\pinocchio~in our case) which each user wants to profile with the library.

\begin{verbatim}
    CC/CXX -I$(PATH_TO_PARALLEL_PMT) -L$(PATH_TO_PARALLEL_PMT) 
    my_code.c/my_code.cpp -lpmt_parallel
\end{verbatim}

\end{document}